\documentclass{aa}

\usepackage{graphicx}
\usepackage{txfonts}
\usepackage{color}
\usepackage{enumitem}
\usepackage{subfigure}
\usepackage{longtable}

\begin{document} 

   \title{ComPRASS: a Combined \textit{Planck}-RASS catalogue of X-ray-SZ clusters}

   \author{P. Tarr\'io
		   	\inst{1}
          \and
          J.-B. Melin
          \inst{2}
          \and
          M. Arnaud
          \inst{1}
          }

   \institute{IRFU, CEA, Université Paris-Saclay, F-91191 Gif-sur-Yvette, France \\
   	Université Paris Diderot, AIM, Sorbonne Paris Cité, CEA, CNRS, F-91191, Gif-sur-Yvette, France\\
   	\email{paula.tarrio-alonso@cea.fr}
   	\and
   	IRFU, CEA, Université Paris-Saclay, F-91191 Gif-sur-Yvette, France\\
   }

\date{Submitted 24 December 2018 / Accepted 28 February 2019}
   
   \abstract{We present the first all-sky catalogue of galaxy clusters and cluster candidates obtained from joint X-ray-SZ detections using observations from the \emph{Planck} satellite and the ROSAT all-sky survey (RASS). The catalogue contains 2323 objects and has been validated by careful cross-identification with previously known clusters. This validation shows that 1597 candidates correspond to already known clusters, 212 coincide with other cluster candidates still to be confirmed, and the remaining 514 are completely new detections. With respect to \textit{Planck} catalogues, the ComPRASS catalogue is simultaneously more pure and more complete. Based on the validation results in the SPT and SDSS footprints, the expected purity of the catalogue is at least 84.5\%, meaning that more than 365 clusters are expected to be found among the new or still to be confirmed candidates with future validation efforts or specific follow-ups.  
   	}


   \authorrunning{P. Tarr\'io et al.}
   \maketitle

\section{Introduction}\label{sec:intro}

Galaxy cluster catalogues with high purity and completeness are fundamental to study clusters both from an astrophysical and a cosmological perspective. Since the first galaxy cluster catalogue constructed by \citet{Abell1958} by analysing photographic plates and which only contained low-redshift clusters, numerous catalogs have been compiled using observational data sets at different wavelengths, from microwaves to X-rays, trying to improve the completeness and purity at all redshift and mass ranges. 

The first cluster catalogues were built from optical data sets, where clusters are identified as overdensities of galaxies. This approach was also used later with infrared data. Clusters can also be detected in X-ray observations, where they appear as bright extended sources. In these images we see the emission of the hot gas of the intracluster medium (ICM). Finally, over the last decade, this gas has also begun to be detected thanks to the characteristic spectral distortion it produces on the cosmic microwave background (CMB) due to Compton scattering of the CMB photons by the ICM electrons. This effect is known as the Sunyaev–Zeldovich (SZ) effect \citep{Sunyaev1970,Sunyaev1972}.

To date, cluster catalogues have been usually constructed through single-survey data, each probing a different region of the electromagnetic spectrum. However, this approach does not leverage the possible synergies that different sources of information might offer if a joint detection approach were used.  
A joint approach would allow to reduce contamination, and thus, to increase the purity of the constructed catalogues, since a contaminant source at a given wavelength will not be generally present at a different wavelength. On the other hand, it would also improve the detection efficiency: an object below the detection threshold of two independent surveys might be detected when combining the two, since its signal will be boosted. 

Although multi-wavelength, multi-survey detection of clusters has been theoretically conceived some years ago \citep{Maturi2007,Pace2008}, it is a very complex task and, until recently, it had only been attempted in the pilot study of \citet{Schuecker2004} on the X-ray ROSAT All-Sky Survey (RASS) \citep{Truemper1993,Voges1999} and the optical Sloan Digital Sky Survey (SDSS, \cite{York2000}). Recognizing the opportunity of exploiting more efficiently existing observations to detect new galaxy clusters, we have recently developed a novel cluster detection method based on the combination of SZ and X-ray surveys \citep{Tarrio2016,Tarrio2018}.  

The method is based on matched multifrequency filters (MMF) \citep{Herranz2002, Melin2006, Melin2012} and does a true joint X-ray--SZ detection by treating the X-ray image as an additional frequency to be simultaneously filtered with the SZ frequency maps. In order to do so, it leverages the expected physical relation between SZ and X-ray fluxes. The X-ray--SZ joint detection method can be seen as an evolution of the MMF3 detection method, one of the MMF methods used to detect clusters from \textit{Planck} observations, that incorporates X-ray observations to improve the detection performance. In \citet{Tarrio2018}, we evaluated this method in the area of the sky covered by the SPT survey using data from the RASS and \textit{Planck} surveys. We showed that, thanks to the addition of the X-ray information, the method is able to simultaneously achieve better purity, better detection efficiency, and better position accuracy than its predecessor, the \textit{Planck} MMF3 detection method. We also showed that if the redshift of a cluster is known by any other means, the joint detection provides a good estimation of its mass.

In this paper, we present ComPRASS, a Combined \textit{Planck}-RASS catalogue of X-ray--SZ sources. This all-sky catalogue contains 2323 galaxy cluster candidates and was constructed by applying the X-ray--SZ joint detection method proposed in \citet{Tarrio2018} on all-sky maps from the \textit{Planck} and RASS surveys. We present an external validation of the catalogue using existing X-ray, SZ and optical cluster catalogues. From this validation, we confirm 1597 ComPRASS candidates. The 726 remaining objects are either new objects (514) or are associated to an already known but yet unconfirmed cluster candidate (212). ComPRASS catalogue is simultaneaously more pure and more complete than \textit{Planck} catalogues, with an expected purity greater than 84.5\% and at least 365 real clusters, unknown to date, among the new or yet-to-confirm candidates.

The structure of the paper is as follows. Section \ref{sec:catalogue} describes the construction of the ComPRASS catalogue, including a brief description of the input data and the detection method. In Sect. \ref{sec:validation} we present the external validation of the catalogue, which is based on cross-identification with previously known clusters and cluster candidates from SZ, X-ray and optical catalogues. Section \ref{sec:evaluation} evaluates some of the properties of the catalogue by comparing it to other catalogues constructed from the same observations. Finally, we conclude the paper and discuss ongoing and future research directions in Sect. \ref{sec:conclusions}. The full catalogue is available in machine readable format. A full description of the available information is given in Appendix A.

Throughout, we adopt a flat $\Lambda$CDM cosmological model with $H_0 = 70$ km s$^{-1}$ Mpc$^{-1}$ and $\Omega_{\rm M} = 1-\Omega_{\Lambda} = 0.3$. We define $R_{500}$ as the radius within which the average density of the cluster is 500 times the critical density of the universe, $\theta_{500}$ as the corresponding angular radius, and $M_{500}$ as the mass enclosed within $R_{500}$.

\section{Construction of the ComPRASS catalogue}\label{sec:catalogue}

The ComPRASS catalogue was obtained by applying the X-ray--SZ joint detection method proposed in \citet{Tarrio2018} to all-sky maps from the \textit{Planck} and RASS surveys. We note that the method of \citet{Tarrio2018} has not been modified for this paper. In rest of this section, we briefly describe the observations that we used, we summarize the X-ray--SZ detection method, indicating the chosen values for various selectable parameters, 
we present the resulting catalogue and we provide an estimation of its completeness as a function of mass and redshift.

\subsection{Input data}\label{ssec:data}

The ROSAT All-Sky Survey (RASS) is the only full-sky X-ray survey performed to date \citep{Truemper1993,Voges1999}. The RASS data release\footnote{ftp://legacy.gsfc.nasa.gov/rosat/data/pspc/processed\_data/rass/release, or http://www.xray.mpe.mpg.de/rosat/survey/rass-3/main/help.html\#ftp} consists of 1378 fields that provide the exposure map and the X-ray counts in three different bands: TOTAL (0.1-2.4 keV), HARD (0.5-2.0 keV), and SOFT (0.1-0.4 keV). Each field covers 6.4 $\deg$ x 6.4 $\deg$ of sky and has a size of 512 x 512 pixels, yielding a resolution of 0.75 arcmin/pixel.

To construct the ComPRASS catalogue, we used an X-ray all-sky HEALPix map that we built from the RASS HARD band information and the RASS exposure maps. This map has a resolution of 0.86 arcmin/pixel, which is HEALPix resolution closest to the RASS resolution. The details of its construction can be found in Appendix B of \citet{Tarrio2016}. This map is available in electronic format.

\textit{Planck} is the only all-sky SZ survey. It observed the sky in nine frequency bands with two instruments: the Low Frequency Instrument (LFI), which covered the 30, 44, and 70 GHz bands, and the High Frequency Instrument (HFI), which covered the 100, 143, 217, 353, 545, and 857 GHz bands. 

To construct the ComPRASS catalogue, we use the six temperature channel maps of HFI, which are the same channels used by the \textit{Planck} Collaboration to produce their cluster catalogues \citep{PlanckEarlyVIII,Planck2013ResXXIX, Planck2015ResXXVII}. In particular, we used the latest version of these maps; their description can be found in \cite{Planck2015ResVIII}. The resolution of the published maps is 1.72 arcmin/pixel. As in \citet{Tarrio2016} and \citet{Tarrio2018}, to make them directly compatible with the all-sky X-ray map mentioned above, we up-sampled them to a resolution of 0.86 arcmin/pixel by zero-padding in the spherical harmonics domain, i.e. by adding new modes with zero power.

\subsection{Joint detection method}\label{ssec:method}
The X-ray--SZ joint detection method proposed in \citet{Tarrio2018} is based on considering the X-ray map as an additional frequency to be simultaneously filtered with the SZ frequency maps. In order to do so, the X-ray map needs to be converted into an equivalent SZ map at a reference frequency $ \nu_{\rm ref} $, leveraging the expected physical relation between SZ and X-ray fluxes, namely the $F_{\rm X}/Y_{500}$ relation. The details of this conversion are described in Appendix B of \citet{Tarrio2016}. The reference frequency $ \nu_{\rm ref} $ is just a fiducial value with no effect on the detection algorithm. In our case, we took $\nu_{\rm ref}$ = 1000 GHz. For the  $F_{\rm X}/Y_{500}$ relation we assumed the relation found by the \citet{PlanckIntI2012}, fixing the redshift to a reference value of $z_{\rm ref}=0.8$, as done in \citet{Tarrio2018}. As shown in the right panel of Fig. 16 of \citet{Tarrio2016}, a change in the normalization of the assumed $F_{\rm X}/Y_{500}$ relation by a factor of 2 only impacts the S/N measurement up to 5\%, which makes the detection robust against possible errors in the assumed relation.

After this conversion, the complete set of maps (the original $N_{\nu}$ SZ maps obtained at sub-mm frequencies $\nu_1, ..., \nu_{N_\nu}$, and the additional map at the reference frequency $ \nu_{\rm ref} $ obtained from the X-ray map) are filtered using the following X-ray--SZ MMF\footnote{We use $\mathbf{k}$ to denote the 2D spatial frequency, corresponding to the 2D position $\mathbf{x}$ in the Fourier space. We use $k$ to denote its modulus. All the variables expressed as a function of $\mathbf{k}$ or $k$ are thus to be understood as variables in the Fourier space.}:
\begin{equation}\label{eq:filter_sz}
\mathbf{\Psi}_{\theta_{\rm s}}(\mathbf{k}) = \sigma_{\theta_{\rm s}}^2 \mathbf{P}^{-1}(k)  \mathbf{F}_{\theta_{\rm s}}(\mathbf{k})
\end{equation}
with
\begin{equation}\label{eq:sigma_sz}
\sigma_{\theta_{\rm s}}^2 =  \left[ \sum_{\mathbf{k}}   \mathbf{F}_{\theta_{\rm s}}^{\rm T}(\mathbf{k})  \mathbf{P}^{-1}(k)  \mathbf{F}_{\theta_{\rm s}}(\mathbf{k}) \right] ^{-1}
,\end{equation}
and
\begin{equation} \label{eq:F_joint}
\mathbf{F}_{\theta_{\rm s}}(\mathbf{k})  = 
[j(\nu_{1}) T_{1}(\mathbf{k}), ..., j(\nu_{N_{\nu}}) T_{N_{\nu}}(\mathbf{k}), C j(\nu_{\rm ref}) T^{\rm{x}}_{\theta_{\rm s}}(\mathbf{k})]^{\rm T}
.\end{equation}
$\mathbf{\Psi}_{\theta_{\rm s}}$ is a $(N_\nu+1) \times 1 $ column vector whose $i$th component will filter the map at observation frequency $\nu_i$;  $\mathbf{P}(k)$ is the noise power spectrum; $j(\nu_{i})$ is the SZ spectral function at frequency $\nu_{i}$;  $T_{i}(\mathbf{k}) = \tilde{T}_{\theta_{\rm s}}(\mathbf{k})  B_{\nu_i}(\mathbf{k})$ and $ T^{\rm{x}}_{\theta_{\rm s}}(\mathbf{k}) = \tilde{T}^{\rm{x}}_{\theta_{\rm s}}(\mathbf{k})  B_{\rm xray}(\mathbf{k})$ are the convolutions of the normalized cluster 2D spatial profiles (SZ and X-ray components, respectively) with the point spread function (PSF) of the instruments at the different frequencies; and the constant $C$ is a geometrical factor that accounts for the different shapes of the SZ and X-ray 3D profiles (Eq. 25 of \citet{Tarrio2016}). A more elaborated description of the filter can be found in \citet{Tarrio2016} and \citet{Tarrio2018}. 

The filtering is done in the Fourier space
as $\hat{y}(\mathbf{k}) = \mathbf{\Psi}^T_{\theta_{\rm s}}(\mathbf{k}) \cdot \mathbf{M}(\mathbf{k})$, where $\mathbf{M}(\mathbf{k}) = [M_{1}(\mathbf{k}),...,M_{N_\nu}(\mathbf{k}),M_{\rm ref}(\mathbf{k})]^T$ is the Fourier transform of the $N_\nu$ +1 input maps. After transforming the result back to the real space, we obtain the filtered map $\hat{y}(\mathbf{x})$ and the S/N map ($\hat{y}(\mathbf{x})$/$\sigma_{\theta_{\rm s}}$). 

We note that the properties of the noise are not the same in different sky regions; therefore, $\mathbf{P}(k)$ has to be calculated locally at each position. This is done in practice from the X-ray and SZ images themselves, assuming that they contain mostly noise. Since this assumption may not hold in some X-ray images due to bright X-ray sources with strong signals, we apply the X-ray source mask defined in \citet{Tarrio2018} for the calculation of $\mathbf{P}(k)$. 

This MMF approach relies on the knowledge of the normalized SZ and X-ray cluster profiles ($\tilde{T}_{\theta_{\rm s}}(\mathbf{k}) $ and $\tilde{T}^{\rm{x}}_{\theta_{\rm s}}(\mathbf{k})$). These profiles are not known in practice, so they need to be approximated by the theoretical profiles that best represent the clusters we want to detect. As in \citet{Tarrio2016} and \citet{Tarrio2018}, we assume the generalized Navarro-Frenk-White (GNFW) profile \citep{Nagai2007} given by 
\begin{equation}\label{eq:pressure_prof}
f(x) \propto \frac{1}{\left( c_{500}x\right) ^{\gamma} \left[ 1+\left( c_{500}x\right) ^{\alpha}\right]^{(\beta-\gamma)/\alpha}  }
\end{equation}
with parameters given by
\begin{equation}\label{eq:sz_param}
\left[ \alpha, \beta, \gamma, c_{500}\right]   = \left[ 1.0510, 5.4905, 0.3081, 1.177\right]
\end{equation}
for the 3D pressure profile \citep{Arnaud2010}, and
\begin{equation}\label{eq:xray_param}
\left[ \alpha, \beta, \gamma, c_{500}\right] = \left[ 2.0, 4.608, 1.05, 1/0.303\right] 
\end{equation}
for the square of the gas density profile \citep{Piffaretti2011}, respectively. We note that $x=r/R_{500}$ represents here the 3D distance to the centre of the cluster in $R_{500}$ units, and $R_{500}$ relates to the characteristic cluster scale $R_{\rm s}$ through the concentration parameter $c_{500}$ ($R_{\rm s} = R_{500}/c_{500}$). The cluster SZ and X-ray profiles as a function of the scaled angular radius ($\theta/\theta_{500}=x$) are then obtained by numerically integrating these 3D GNFW profiles along the line of sight. 

Finally, these normalized cluster profiles need to be convolved by the instrument beams ($B_{\nu_i}(\mathbf{k})$ and $B_{\rm xray}(\mathbf{k})$). For the SZ components, we use a Gaussian PSF with FWHM depending on the frequency, as shown in Table 6 of \citet{Planck2015ResVIII}. For the X-ray component, we use a PSF that was estimated numerically by stacking observations of X-ray point sources from the Bright Source Catalogue \citep{Voges1999}. These are the same instrument beams used in \citet{Tarrio2016} and \citet{Tarrio2018}, since we have the same input observations.

To implement the detection procedure in practice, we proceed in two phases.
\begin{enumerate}
	\item First phase: we project the all-sky maps into 504 small $10\degr \times 10\degr$ tangential patches, as was done in MMF3 \citep{PlanckEarlyVIII,Planck2013ResXXIX,Planck2015ResXXVII}. Each patch is filtered by the X-ray--SZ filter $\mathbf{\Psi}_{\theta_{\rm s}}$ (Eq. \ref{eq:filter_sz}) using $N_{\rm s}=32$ different sizes, covering the expected range of radii. In our case, we vary $\theta_{500}$ from 0.94 to 35.31 arcmin, in $N_{\rm s}=32$ steps equally spaced in logarithmic scale. For each size, we obtain a filtered map and a S/N map. Then, we construct a list with the peaks in these maps that are above a specified S/N threshold $q$. The iterative  procedure for finding the peaks in each set of $N_{\rm s}$ S/N maps is described in \citet{Tarrio2018}. Finally, we merge the 504 lists into a single preliminary all-sky list of candidates by merging peaks that are close to each other by less than 10 arcmin, as was done in MMF3 \citep{PlanckEarlyVIII,Planck2013ResXXIX,Planck2015ResXXVII}. 
	\item Second phase: we create a set of maps centered at each candidate from the first phase, and we re-apply the MMF (see details of the procedure in \citet{Tarrio2018}). This second phase allows a better estimation of the candidate properties and S/N. Only the candidates whose new S/N is above the specified threshold $q$ are kept in the final list. 
\end{enumerate}

A key point of this X-ray--SZ detection algorithm is the selection of the threshold $q$ that defines which S/N peaks are kept in the first and second phases. As explained in \citet{Tarrio2018}, the X-ray--SZ method uses an adaptive threshold that depends on the noise characteristics of each region, and that is determined numerically by means of Monte Carlo simulations. These simulations establish, as a function of the mean Poisson noise level in the X-ray map, the mean Gaussian noise level in the SZ maps and the filter size ($\theta_{\rm s}$), the joint S/N threshold $q_{\rm J}$ corresponding to a given false alarm probability $P_{\rm{FA}}$ (i.e. the probability that a detection is due to a random fluctuation). The value of $P_{\rm{FA}}$ serves to select the operational point of the detection method, with higher values resulting in more complete but less pure catalogues and lower values yielding more pure but less complete catalogues. To construct the ComPRASS catalogue, we used $P_{\rm{FA}}=3.4 \cdot 10^{-6}$, which corresponds to a cut at $4.5\sigma$ in a zero-mean Gaussian distribution.
For reasons of simplicity, the adaptive threshold $q_{\rm J}$ is applied after obtaining the final candidate list from the second phase. The threshold $q$ to be applied in the first and second phases is set to $q = 4$, a sufficiently low value so that it does not introduce any different selection effect, i.e. it does not discard any candidate above $q_{\rm J}$.

\begin{figure}[]
	\centering
	\includegraphics[width=0.99\columnwidth]{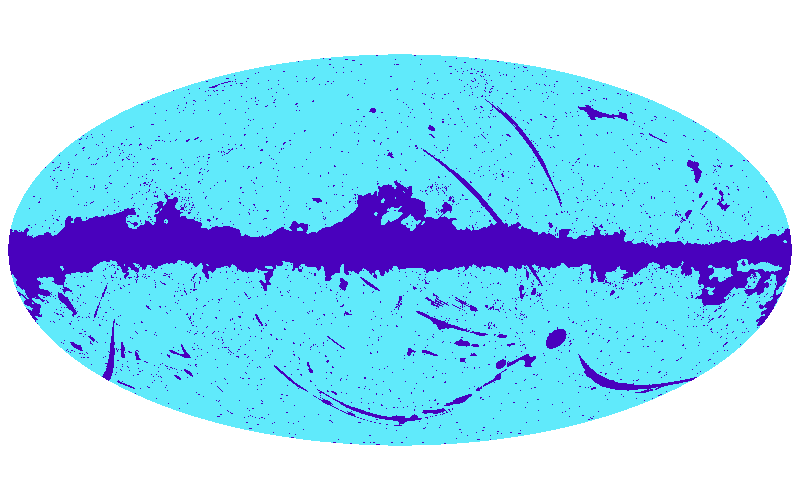}
	\caption{Mask applied to the second phase detections. It results from the combination of the SZ mask and the X-ray exposure time mask. The dark blue areas are masked, the light blue area is the final footprint, which accounts for 80.96\% of the sky. The electronic version of this mask is available.}
	\label{fig:mask}
\end{figure}

Finally, the catalogue produced at the second phase needs to be further cleaned to discard detections in contaminated regions of the sky, in regions with poor statistics, or that correspond to non-cluster objects (mainly AGNs). To this end, we apply an SZ mask to avoid SZ contaminated regions and a X-ray exposure time mask ($t_{\rm exp}>100$ s) to avoid X-ray regions with poor statistics. These two masks are defined in Sect. 3.4.1 and 3.4.2 of \citet{Tarrio2018}, respectively. Fig. \ref{fig:mask} shows the combination of these two masks, which provide a final footprint of 80.96\% of the sky. To discard AGN detections, we establish an additional criterion based on the SZ part of the S/N: we only keep the detections satisfying (S/N)$_{\rm SZ}$ > 3, as described in Sect. 3.4.3 of \citet{Tarrio2018}.

\subsection{ComPRASS catalogue}\label{ssec:catalog}

\begin{figure}[]
	\centering
	\subfigure[]{\includegraphics[width=0.99\columnwidth]{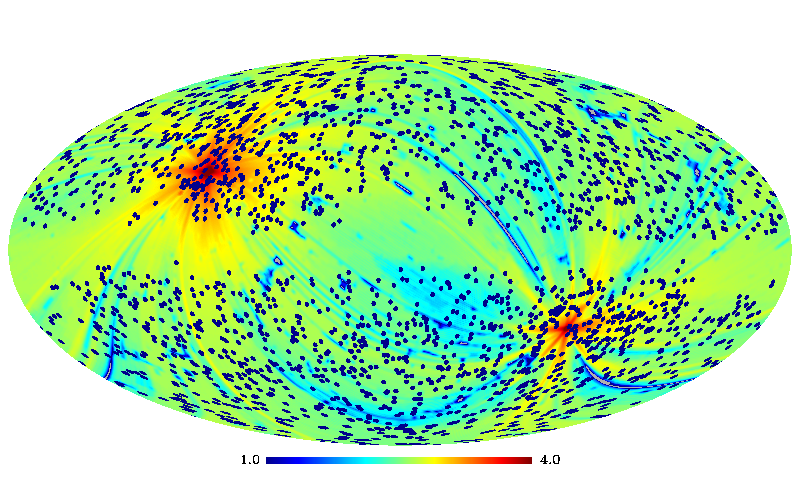}}
		\subfigure[]{\includegraphics[width=0.99\columnwidth]{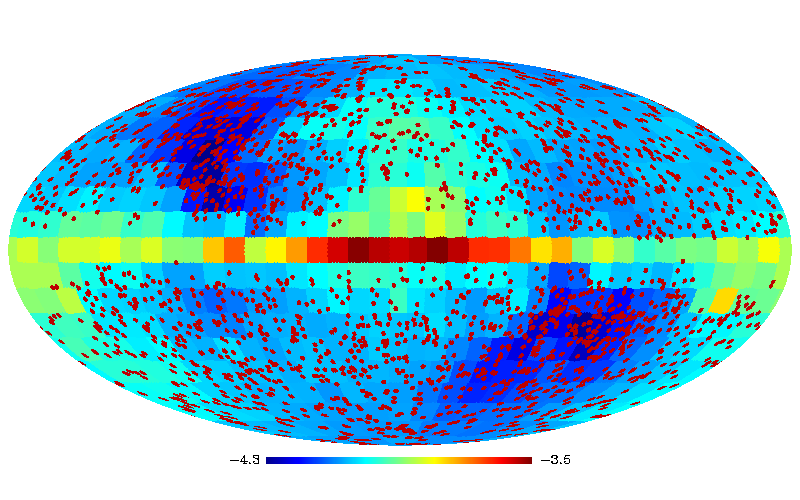}}
	\caption{Sky distribution of the 2323 ComPRASS candidates (blue/red dots). The sky map is colour-coded according to (a) the logarithm of the RASS exposure time and (b) the logarithm of the \textit{Planck} noise standard deviation map.}
	\label{fig:detections_in_sky_map}
\end{figure}

The ComPRASS catalogue was obtained by running the blind joint X-ray--SZ detection algorithm summarized in Sect. \ref{ssec:method} on the RASS and \textit{Planck} all-sky maps described in Sect. \ref{ssec:data}. It contains 2323 candidates, distributed in the sky as shown in Fig. \ref{fig:detections_in_sky_map}. 
The sky is not covered homogeneously: there are more candidates in the regions where the RASS exposure time is higher and where the \textit{Planck} noise is lower. This is expected, since in those regions both surveys are deeper. 

\real{Table \ref{table:catalogue}} summarizes the main properties of the ComPRASS candidates.  
For each candidate, the catalogue provides its position, the joint S/N: (S/N)$_{\rm J}$, and the SZ and X-ray components of this S/N: (S/N)$_{\rm SZ}$ and (S/N)$_{\rm XR}$. (S/N)$_{\rm J}$ is the value of the joint S/N map ($\hat{y}(\mathbf{x})$/$\sigma_{\theta_{\rm s}}$) at the position and size of the detection. (S/N)$_{\rm SZ}$ ((S/N)$_{\rm XR}$) corresponds to the S/N of the SZ (X-ray) filtered maps, i.e. the SZ (X-ray) filtered maps divided by the background noise of the SZ (X-ray) maps, at the position and size of the blind joint detection. 
It also provides a value for the significance of each detection. This value is defined in \citet{Tarrio2018} as the significance value in a Gaussian distribution corresponding to the probability that the detection is due to noise, which is calculated from the results of the Monte Carlo simulations performed to calculate the joint S/N threshold $q_{\rm J}$. 

The electronic version of the catalogue also contains the mass-redshift degeneracy curves for each candidate. These curves provide the mass ($M^{\rm XSZ}_{500}$) estimated by the joint X-ray--SZ extraction for various values of redshift between 0.01 and 1.2. They are obtained by 1) re-extracting the candidate signal at the position given by the blind detection using different values for the reference redshift $z_{\rm ref}$ (0.01, 0.1, 0.2, 0.4, 0.7 and 1.2); 2) interpolating between the six resulting $Y_{500}(\theta_{500})$ degeneracy curves to obtain the curve corresponding to a given value of redshift between 0.01 and 1.2; and 3) breaking the size–flux degeneracy using the $M_{500} - D^2_A Y_{500}$ scaling relation from \citet{Planck2013ResXX}: 
\begin{equation}\label{eq:M500-Y500}
E^{-2/3}(z)\left[\frac{D^2_A(z)Y_{500}}{10^{-4}\rm{Mpc}^2}\right] = 10^{-0.19}\left[\frac{M_{500}}{6\cdot 10^{14}M_\odot}\right]^{1.79},
\end{equation}
which relates $\theta_{500}$ and $Y_{500}$ when $z$ is known, as explained in  \cite{Planck2013ResXXIX} and \citet{Tarrio2018}. This relation does not take into account possible bias between X-ray derived mass and true mass. The upper and lower bounds of the 68\% confidence interval in the mass-redshift degeneracy curves are also provided in the electronic version of the catalogue. These bounds are derived from the bounds in the $Y_{500}(\theta_{500})$ curves, using the procedure explained above. The errors on the $Y_{500}(\theta_{500})$ curves are calculated from Eq. 32 of \citet{Tarrio2016}. This expression corrects the error on $Y_{500}$ obtained by the filter (Eq. 28 of \citet{Tarrio2016}) by a factor that depends on the size $\theta_{500}$ and that accounts for the additional dispersion produced by the mismatch between the real cluster profiles and the ones used in the filter. We note that the error on $Y_{500}$ obtained by the filter (Eq. 28 of \citet{Tarrio2016}) does not only depend on the S/N, but also on the variance of the filtered Poisson fluctuations on the signal.

The catalogue also contains the information obtained from the external validation (see details in Sect. \ref{sec:validation}). If the candidate is associated with a previously known cluster or cluster candidate, the name of the associated object is indicated. Moreover, if the redshift of the associated cluster is known, it is also included in the ComPRASS catalogue. In the case of multiple associations with different redshifts, the order of priority defined in Sect. \ref{ssec:redshift_priority} is applied. When the redshift of the cluster candidate is known from the external validation, the catalogue provides an estimation of its mass ($M_{\rm J}$), and the upper and lower bounds of its 68\% confidence interval, calculated from the mass-z degeneracy curves at the redshift of the cluster. 

A full description of all the fields provided in the ComPRASS catalogue is given in Appendix \ref{app:description}.

\subsection{Completeness estimation}\label{ssec:selection_function}

In this section we provide an estimation of the completeness to illustrate the performance of the ComPRASS catalogue at different mass and redshift ranges. 

\begin{figure}[]
	\centering
	\includegraphics[width=0.99\columnwidth]{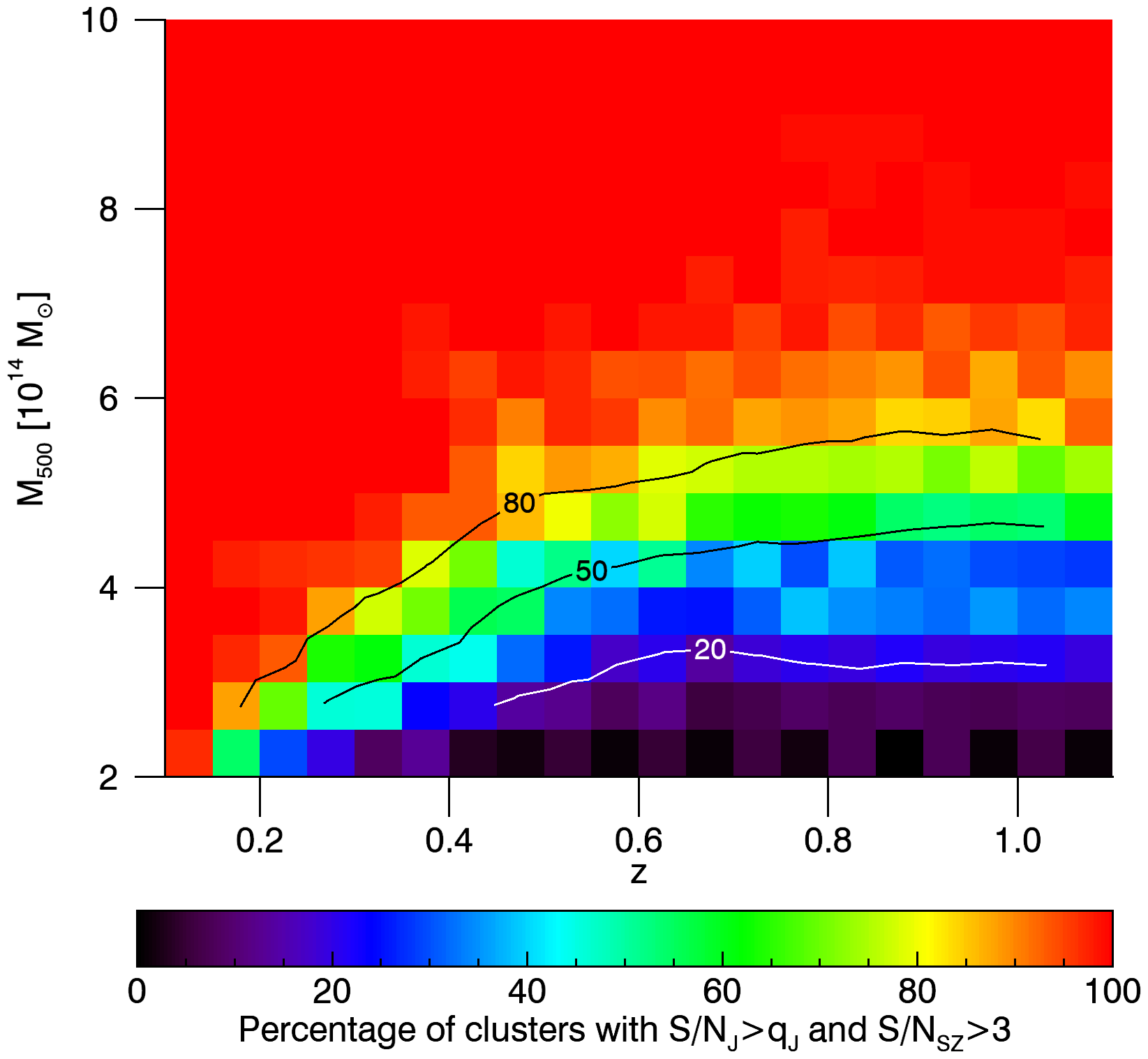}
	\caption{Estimated completeness of the ComPRASS catalogue in different mass-redshift bins.}
	\label{fig:completeness}
\end{figure}

The completeness has been calculated by injecting simulated clusters into the real RASS and \textit{Planck} maps described in Sect. \ref{ssec:data} and extracting them with the filter described in Sect. \ref{ssec:method}. 
The clusters whose extracted S/N satisfy the thresholds imposed in the blind detection (S/N$_{\rm J} > q_J$ and S/N$_{\rm SZ} > 3$), are considered to be detected. The completeness is then calculated as the fraction of detected clusters with respect to the total number of clusters injected outside the masked region of the sky (see Fig.\ref{fig:mask}). For simplicity, we fixed the positions, sizes and redshifts of the clusters for the extraction step. Fig. \ref{fig:completeness} shows the resulting completeness as a function of mass and redshift. For a mass of $5\cdot10^{14} M_\odot$, the catalogue should contain ~80\% of the clusters at z=0.5 and ~60\% of the clusters at z=1.

For these simulations, we considered four mass bins: 2-4, 4-6, 6-8, and 8-10 $\cdot 10^{14} M_{\odot}$ and eight redshift bins: 0.1-0.3, 0.3-0.5, 0.5-0.6, 0.6-0.7, 0.7-0.8, 0.8-0.9, 0.9-1.0, and 1.0-1.1, resulting in 32 bins in the redshift-mass plane. For each bin, we injected 1000 clusters at random positions of the sky, with $z$ and $M_{500}$ uniformly distributed in the bin. The resulting 32000 clusters were then re-binned into smaller mass-redshift bins for visualisation purposes in Fig. \ref{fig:completeness}. The clusters were simulated as in \citet{Tarrio2016} (Sect. 2.2.1 and Sect. 4.2.1), assuming the GNFW profile defined in Eq. \ref{eq:pressure_prof} with the parameters given by Eq. \ref{eq:sz_param} and Eq. \ref{eq:xray_param}, for the SZ and X-ray components, respectively. The amplitudes of these simulated clusters are estimated according to the expected values of $Y_{500}$ (for SZ) and $L_{500}$ (for X-ray). $L_{500}$ and the corresponding flux $F_{\rm X}$ were calculated from the L-M relation in \cite{Arnaud2010,PlanckEarlyXI}, including the scatter $\sigma_{\rm{log} L}=0.183$. $Y_{500}$ was calculated from the nominal flux $F_{\rm X}$ assuming the $F_{\rm X}/Y_{500}$ relation found by the \citet{PlanckIntI2012}, implicitly neglecting the scatter of the Y-M relation. 

We note that this selection function should not be used for precise cosmological purposes. A precise selection function would require a full autoconsistent modelling of: the scatter in the Y-M relation, the deviations from the assumed cluster profile, the uncertainty on the redshift at the detection step, etc. as a function of the position on the sky.

\section{External validation}\label{sec:validation}
The ComPRASS catalogue is validated by identifying the candidates that are associated with previously known clusters. To this end, we use existing SZ, X-ray and optical cluster catalogues. 

In particular, we took several SZ-selected catalogues, namely the three \textit{Planck} catalogues: ESZ \citep{PlanckEarlyVIII}, PSZ1 \citep{Planck2013ResXXIX, Planck2013ResXXXII}, and PSZ2 \citep{Planck2015ResXXVII}, the SPT catalogue \citep{Bleem2015}, and the ACT and ACTPol catalogues \citep{Hasselfield2013,Hilton2018}. The redshifts in the original PSZ1 and PSZ2 catalogues were updated with follow-up data from MegaCam at CFHT \citep{vanderBurg2016}, NOT (Dahle et al., in prep.), WHT \citep{Buddendiek2015}, the PSZ1 follow-up program developed at Roque de los Muchachos Observatory \citep{Barrena2018}\footnote{We only updated the clusters with definitely confirmed optical counterparts (validation flag = 1) and those with no counterpart found (validation flag = ND).}, the high-z PSZ2 follow up program of \cite{Burenin2018}, the optical follow-up of PSZ2 clusters of \cite{Boada2018}, and from some private communications on specific clusters (see notes in ComPRASS catalogue). We also consider the extension of the \textit{Planck} catalogue presented in \cite{Burenin2017}. 
 
We also took as reference the X-ray selected MCXC catalogue \citep{Piffaretti2011}. This is a metacatalogue of X-ray detected clusters that was constructed from publicly available cluster catalogues of two kinds: RASS-based catalogues, obtained from the RASS survey data, and serendipitous catalogues, based on deeper pointed X-ray observations. It contains only clusters with an available redshift at that date. We used an updated version of the catalogue (Sadibekova et al. in prep.), which includes the 1743 original clusters plus 125 new clusters from 160SD, NORAS, WARPS, SGP, REFLEXII \citep{Chon2012} and MACS (including the ones in \cite{Mann2012} and \cite{Repp2018}). This new version takes into account the new spectroscopic redshifts published in the literature for the ROSAT clusters. We furthermore supplement the updated MCXC with 44 unpublished MACS clusters observed by XMM-Newton or Chandra.

In addition, we considered several optically selected cluster catalogues: the Abell catalogue \citep{Abell1989}, the Zwicky catalogue \citep{Zwicky1961}, and five catalogues based on SDSS data, namely, the redMaPPer catalogue \citep{Rykoff2014}, the MaxBCG catalogue \citep{Koester2007}, the GMBCG catalogue \citep{Hao2010}, the AMF catalogue \citep{Szabo2011}, and the WHL catalogue \citep{Wen2012}.

Finally, we also took into account three recently published high-redshift clusters catalogues (\cite{Buddendiek2015}, \cite{Gonzalez2018} and \cite{Wen2018}) and we did a search in NED and SIMBAD databases for possible missing associations.

\begin{figure*}[]
	\centering
	\subfigure[]{\includegraphics[width=0.99\columnwidth]{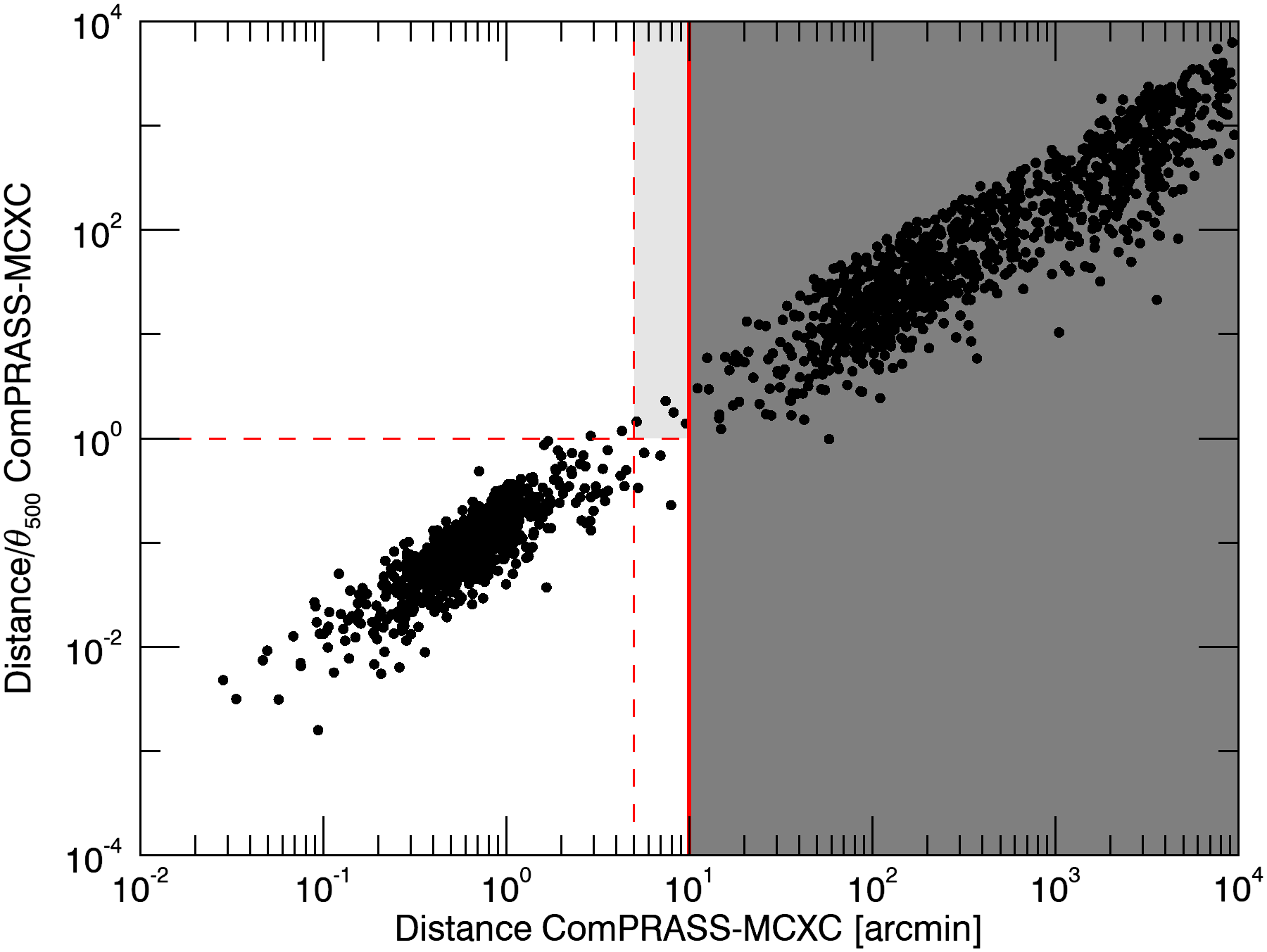}\label{fig:matching_XR_loglog}}	\subfigure[]{\includegraphics[width=0.99\columnwidth]{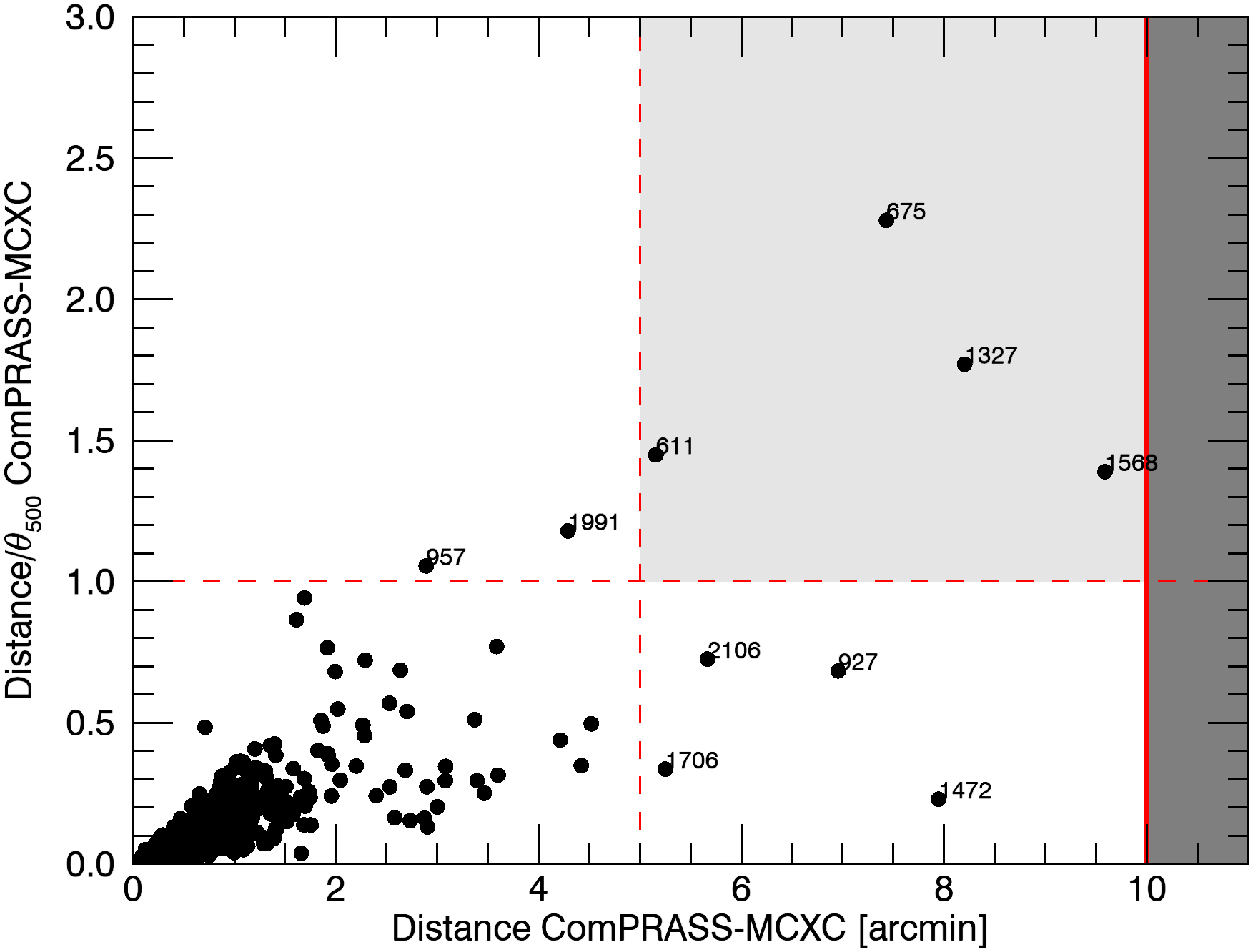}\label{fig:matching_XR_zoom}}
	\caption{Positional criteria for matching ComPRASS candidates to X-ray detected MCXC clusters. The distance between each ComPRASS candidate and its closest MCXC cluster is plotted against their relative distance in terms of $\theta_{500}$. The white areas define the regions where the associations are considered to be correct. (a) shows a complete view in logarithmic scale, (b) shows a zoom of the low-distance region.}
	\label{fig:matching_XR}              
\end{figure*}

Based on the identification with clusters or candidates of these catalogues, the ComPRASS candidates are classified into three classes: 
\begin{itemize}
	\item \textit{Confirmed} (class 1): Candidates that are associated to an already confirmed cluster.
	\item \textit{Identified not confirmed} (class 2): Candidates that are associated to candidates in X-ray or SZ catalogues that have not been yet confirmed.
	\item \textit{New} (class 3): Candidates that are not associated to any cluster or cluster candidate.
\end{itemize}
In the rest of this Section, we describe in detail this external validation.

\subsection{Identification with X-ray clusters}\label{ssec:xmatch_XR}

To identify the ComPRASS candidates that correspond to already known X-ray clusters, we first determined the closest MCXC cluster for each ComPRASS candidate. Then, we relied on two quantities to do the association: the angular distance between the candidate and the closest cluster ($d$), and their relative distance in terms of the size of the cluster ($d/\theta_{500}$), calculated from the mass and redshift reported in the MCXC catalogue. Figure \ref{fig:matching_XR_loglog} shows a scatter plot of the angular distance versus the relative distance between all the ComPRASS candidates and their closest MCXC cluster. We observe two main clouds of points: those with a small distance in absolute and in relative terms, and those with a long distance in absolute and in relative terms. The first cloud of points represents good associations, whereas the second cloud corresponds to the candidates that are randomly distributed with respect to the considered known clusters. There are also a few points between the two main clouds (see zoom in Fig. \ref{fig:matching_XR_zoom}) for which the association is not clear. These intermediate cases were studied individually (see Appendix \ref{app:notes} for details) to conclude that only four of them did not correspond to a true association: candidates 611, 675, 1327, and 1568. From this analysis, we decided to use the following association rules:

\begin{itemize}
	\item If $d>10'$ the candidate is not associated with the cluster (dark grey area in Fig. \ref{fig:matching_XR}).
	\item If $5'<d<10'$ and $d/\theta_{500}>1$ the candidate is not associated with the cluster (light grey area in Fig. \ref{fig:matching_XR}).
	\item Otherwise ($5'<d<10'$ and $d/\theta_{500}<1$, or $d<5'$) the candidate is associated with the cluster (white area in Fig. \ref{fig:matching_XR}).
\end{itemize}

These association rules give us 737 matches of ComPRASS candidates with MCXC clusters. 696 candidates are associated with MCXC clusters with known redshift and mass. The remaining 41 are associated with MCXC clusters for which we do not know $\theta_{500}$ (7 clusters without redshift and 34 MACS clusters without a published $L_{500}$, and thus, without $M_{500}$). Although we cannot calculate the relative distance for these 41 associations, all of them have $d<3$ arcmin, so they are considered to be correct. 730 of these 737 associations are classified as confirmed (class 1), while the remaining 7, corresponding to the 7 clusters without redshift, are classified as identified but not confirmed (class 2).

\begin{figure}[]
	\centering
	\includegraphics[width=0.99\columnwidth]{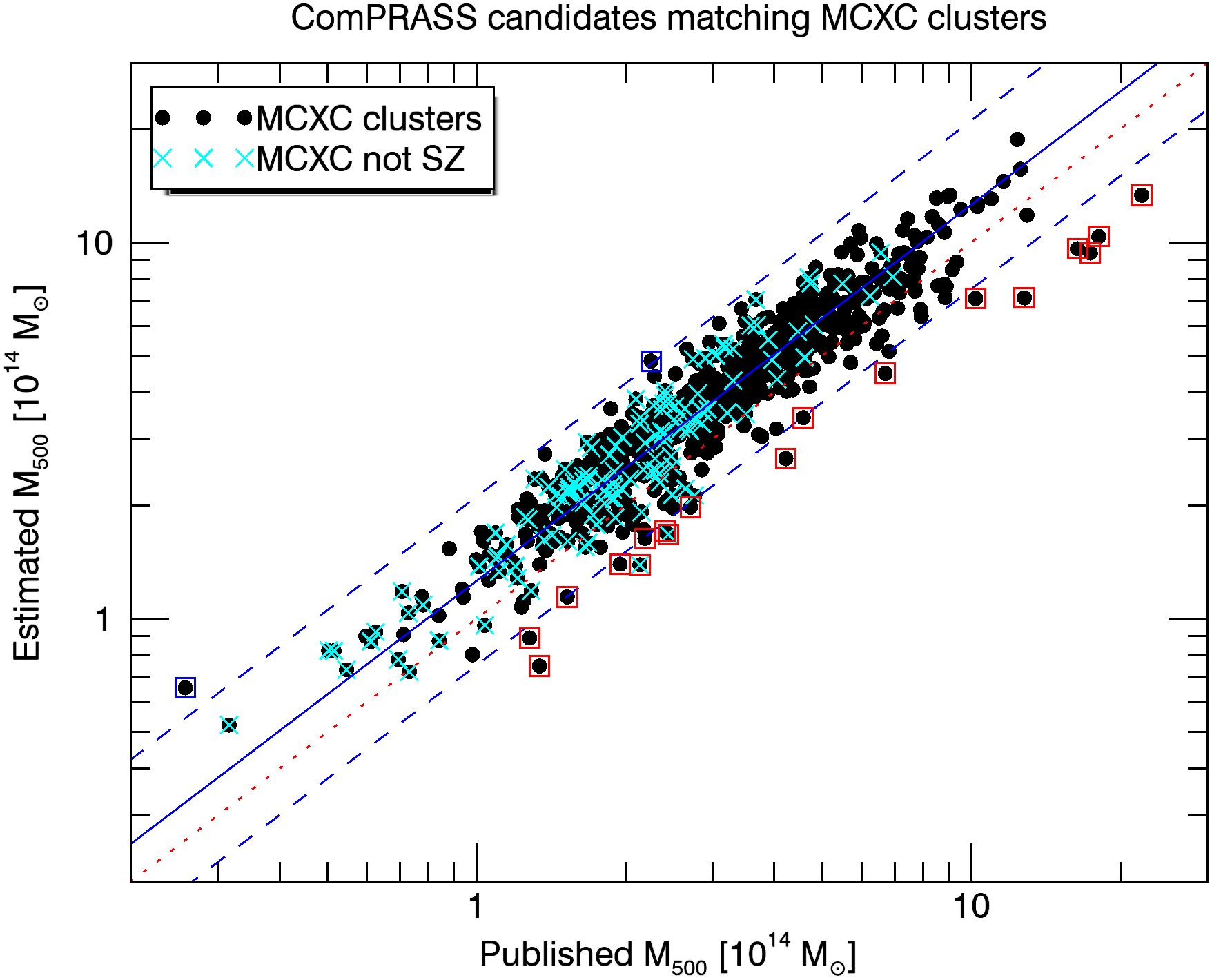}
	\caption{$M_{500}$ estimated from the joint detection for the 696 ComPRASS candidates matching a confirmed MCXC cluster vs. the published $M_{500}$ of the corresponding cluster. The dotted red line indicates the line of zero intercept and unity slope. The solid blue line indicates the median ratio. The dashed blue lines indicate the interval of $\pm 2.5\sigma$ around the median ratio. Outliers are highlighted with a blue (high estimated mass) or red (low estimated mass) square; see text. Cyan crosses indicate MCXC clusters that are not in \textit{Planck} or SPT. }
	\label{fig:mass_xsz-mass_x}              
\end{figure}

\begin{figure*}[]
	\centering
	\subfigure[]{\includegraphics[width=0.99\columnwidth]{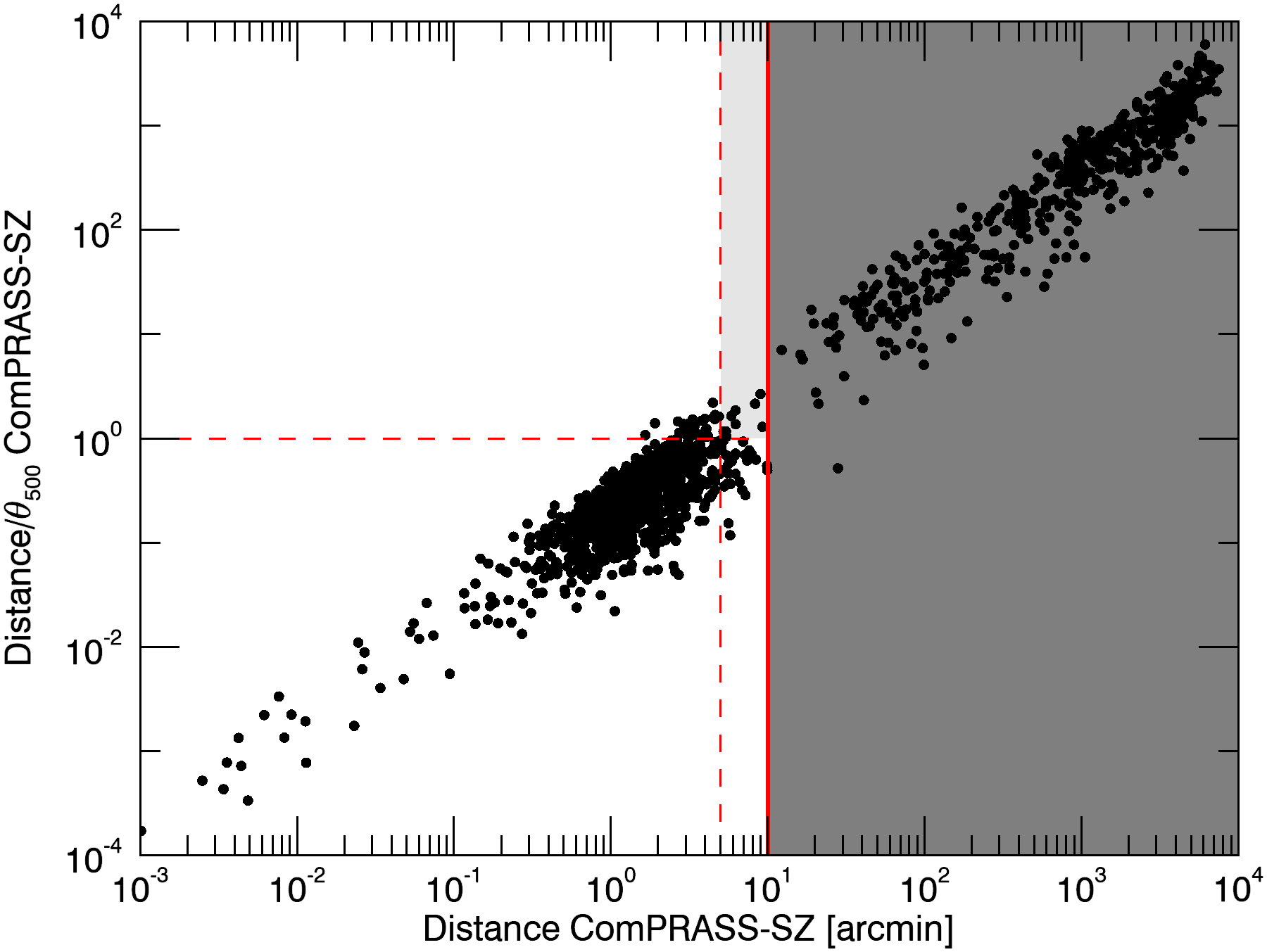}\label{fig:matching_SZ_loglog}}
	\subfigure[]{\includegraphics[width=0.99\columnwidth]{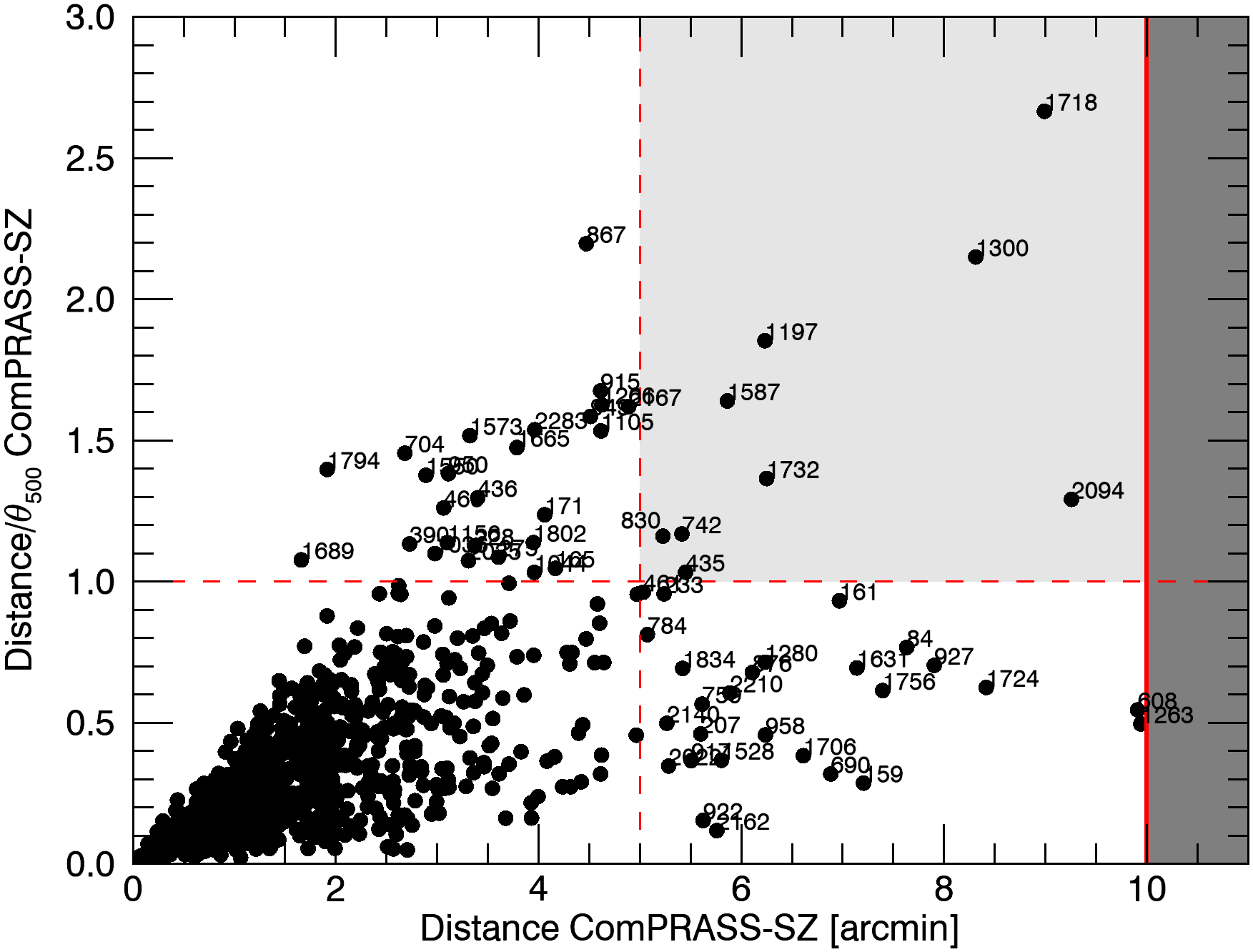}\label{fig:matching_SZ_zoom}}
	\caption{Positional criteria for matching ComPRASS candidates to SZ clusters.
The distance between each ComPRASS candidate and its closest SZ cluster is plotted against their relative distance in terms of $\theta_{500}$. The white areas define the regions where the associations are considered to be correct. (a) shows a complete view in logarithmic scale, (b) shows a zoom of the low-distance region.}
	\label{fig:matching_SZ}              
\end{figure*}

Following the procedure described in Sect. \ref{ssec:catalog} and using the $M_{500} - D^2_A Y_{500}$ relation proposed in \cite{Planck2013ResXX} (see eq. \ref{eq:M500-Y500}), we estimated the mass $M_{500}$ for the 696 detections matching confirmed MCXC clusters with published values for their mass.  Figure \ref{fig:mass_xsz-mass_x} shows the relation between the estimated mass and the published mass for the corresponding clusters.

This mass comparison shows that the ratio between the estimated mass and the published mass is on average slightly greater than one, with a median value of 1.26. The same value is found for the ratio between the published SZ mass and the published MCXC mass for the same clusters. This  behaviour was also observed by the \cite{Planck2015ResXXVII} when they compared the SZ mass and the X-ray luminosity of common PSZ2-MCXC objects. We identified twenty outliers that are at more than 2.5$\sigma$ from the median ratio: two with overestimated mass and eighteen with underestimated mass.

The two ComPRASS candidates with overestimated mass (1490 and 2195) match also PSZ2 clusters with known redshift and mass. The published masses for the corresponding PSZ2 clusters are higher than those of the associated MCXC clusters, and compatible with the joint mass estimates. The redshift of the MCXC and the PSZ2 clusters coincide, so we can conclude that these outliers are due to a difference between the X-ray and the SZ mass estimates, maybe because they are X-ray under-luminous clusters.

Regarding the eighteen candidates with underestimated mass, 10 of them are very extended and present an offset between the SZ and the X-ray peaks, which results in an underestimated mass. The remaining 8 match PSZ2 clusters with published SZ masses which are smaller than the MCXC masses. They are probably clusters with a stronger X-ray emission than the average for a given mass (2 of them are classified as cool-core clusters in \cite{Rossetti2017}), yielding a higher X-ray mass estimate.

This discussion is in agreement with results from hydrodynamical simulations, which indicate that the SZ signal is more tightly correlated to the mass than the X-ray flux. We cannot exclude that some outliers are pathological cases (e.g. very strong mergers) where the SZ signal is very different from what is expected from the true mass.

\subsection{Identification with SZ clusters}\label{ssec:xmatch_SZ}

To identify the ComPRASS candidates that correspond to already known SZ clusters, we followed a procedure analogous to the one described in Section \ref{ssec:xmatch_XR} for the association with MCXC clusters. To this end, we selected only the objects in the considered SZ catalogues (excluding \cite{Burenin2017}) with known redshift and mass (i.e. confirmed clusters). Figure \ref{fig:matching_SZ_loglog} shows a scatter plot of the angular distance versus the relative distance between all the ComPRASS candidates and their closest SZ confirmed cluster (with known redshift and mass). As in the association with MCXC clusters, we observe two main clouds of points: those corresponding to good associations and those corresponding to bad associations. There are also a few points between the two main clouds (see zoom in Fig. \ref{fig:matching_SZ_zoom}) for which the association is not clear. These intermediate cases were studied individually to conclude that only four of them did not correspond to a true association: candidates 1300, 1587, 1718, and 2094 (see Appendix \ref{app:notes} for details). From this analysis, we decided to discard the associations with $d>5'$ and $d/\theta_{500}>1$, as in the association with MCXC clusters. These association rule gives us 1060 matches of ComPRASS candidates with SZ clusters with known redshift and mass. These candidates are considered to be confirmed (class 1). 

Apart from confirmed SZ clusters, the considered SZ catalogues also contain cluster candidates that have not been yet confirmed to date. The redshift of these objects is not known, thus we cannot calculate $\theta_{500}$ and the relative distance for the possible associations. Therefore, we need to define a different association rule for these cases. To this end, we did a positional matching within a 10' search radius and we found 239 candidates matching a not-confirmed SZ candidate. Most of these associations (233) have $d<5$ arcmin, so based on the above study they are considered to be correct. The remaining six (candidates 153, 809, 1487, 1904, 1934, 2152) have $d>5$ arcmin and were analyzed individually. Based on their X-ray and SZ S/N maps, it seemed that most of these associations were correct, so we decided to use the following rule for the associations with SZ candidates: If $d<10'$ the ComPRASS candidate is associated with the SZ candidate; otherwise it is not associated. The candidates associated in this way to an SZ candidate are classified in class 2 (identified not confirmed). If they were already confirmed by the X-ray identification, they remain in class 1 (3 out of 239).

\begin{figure}[]
	\centering
	\includegraphics[width=0.99\columnwidth]{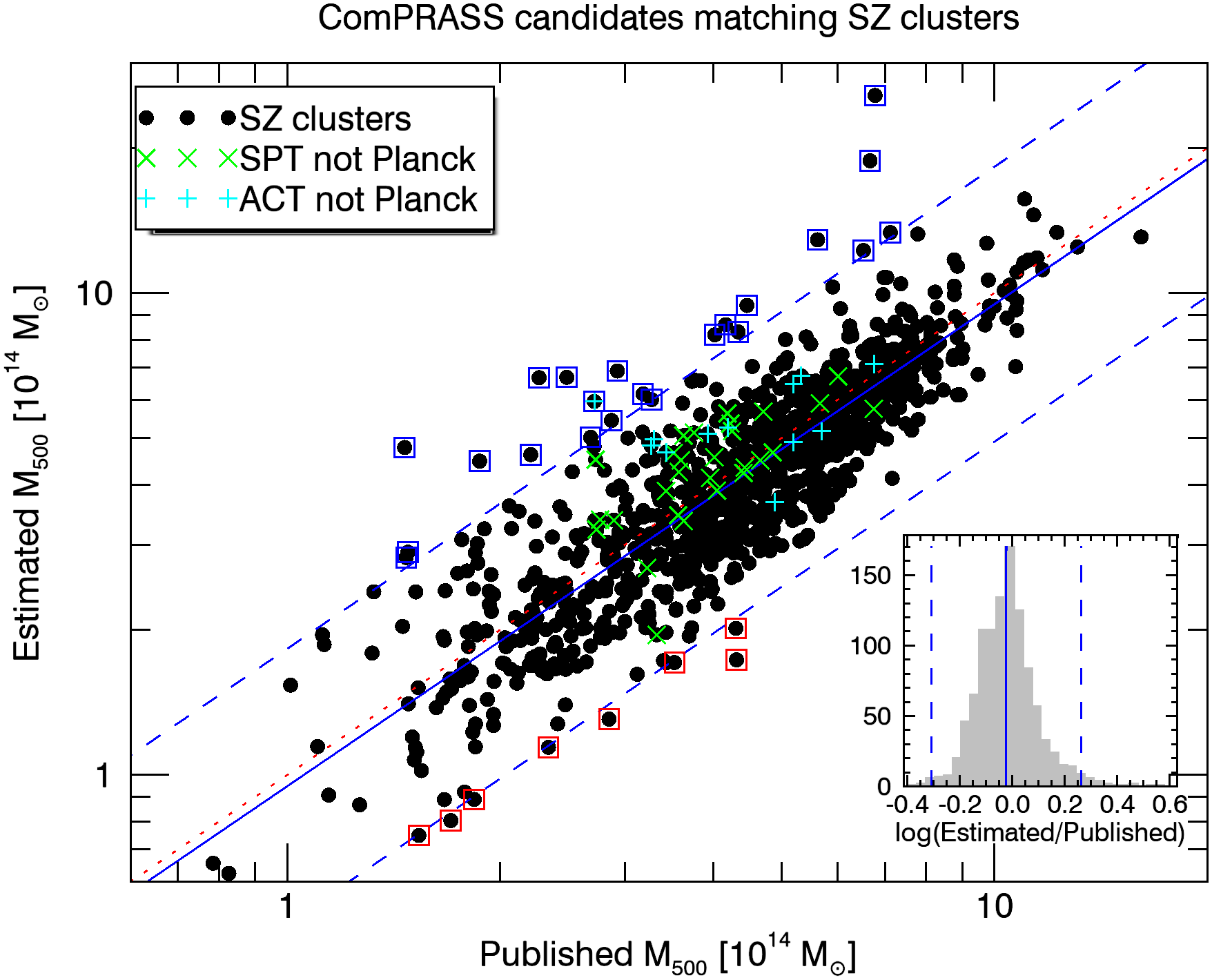}
	\caption{$M_{500}$ estimated from the joint detection for the 1060 ComPRASS candidates matching a confirmed SZ cluster vs. the published $M_{500}$ of the corresponding cluster. The dotted red line indicates the line of zero intercept and unity slope. The solid blue line indicates the median ratio. The dashed blue lines indicate the interval of $\pm 2.5\sigma$ around the median ratio. Outliers are highlighted with a blue (high estimated mass) or red (low estimated mass) square; see text. Green crosses indicate SPT clusters that are not in \textit{Planck}. Cyan crosses indicate ACT clusters that are not in \textit{Planck}. The subpanel shows the histogram of the logarithm of the ratio between the estimated and the published $M_{500}$. }
	\label{fig:mass_xsz-mass_sz}
\end{figure}

We estimated the mass $M_{500}$ for the 1060 detections matching confirmed SZ clusters. Figure \ref{fig:mass_xsz-mass_sz} shows the relation between the estimated mass and the published mass for the corresponding clusters. This mass comparison shows that the ratio between the estimated mass and the published mass is very close to one, with a median value of 0.95.  We identified 30 outliers that are at more than 2.5$\sigma$ from the median ratio: 22 with overestimated mass and 8 with underestimated mass. 

Regarding the 22 ComPRASS candidates with overestimated mass, we found that 18 of them are cool-core clusters, according to \cite{Rossetti2017}, \cite{Andrade2017}, \cite{Ebeling2010}, \cite{Pratt2009}, \cite{Hudson2010} and \cite{Morandi2015}. Other 2 candidates are also cool-core clusters according to I. Bartalucci (priv. comm.) and T. Reiprich and G. Schellenberger (priv. comm.). I. Bartalucci analyzed the XMM observation of  
candidate 1053 and obtained a central density at $0.01 R_{500}$ of 0.058, higher than the limit of 0.015 considered in \cite{Hudson2010}, indicating that it is a cool-core. 
Candidate 2019 is also a cool-core cluster according to the temperature profile obtained from Chandra observations by T. Reiprich and G. Schellenberger (priv. comm.).
Since our mass estimate relies both on the X-ray and SZ signal, and the X-ray emission for these clusters is stronger than the one expected from the SZ emission, our joint mass estimate is higher than the mass estimated from the SZ signal only, which is very close to the published SZ mass. This explains why our joint mass estimates are higher than the published SZ masses for these cool-core clusters. Regarding the remaining 2 ComPRASS candidates with overestimated mass, one is not a cool-core according to \cite{Andrade2017}, but appears as a weak cool-core in \cite{Hudson2010}; and finally for the last one we do not know whether it is a cool-core cluster or not.

\begin{figure*}[]
	\centering
	\subfigure[]{\includegraphics[width=0.99\columnwidth]{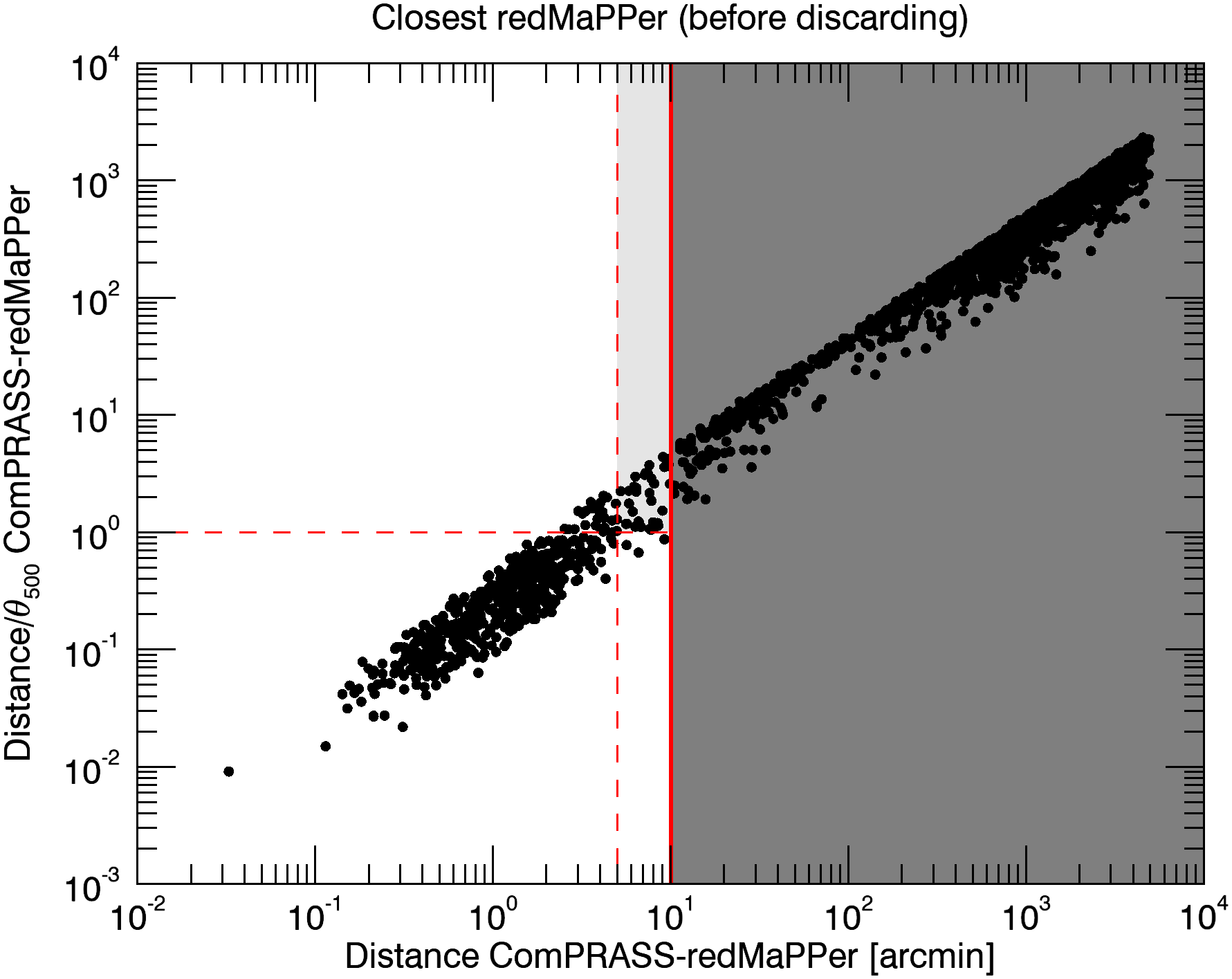}\label{fig:matching_redm_loglog}}
	\subfigure[]{\includegraphics[width=0.99\columnwidth]{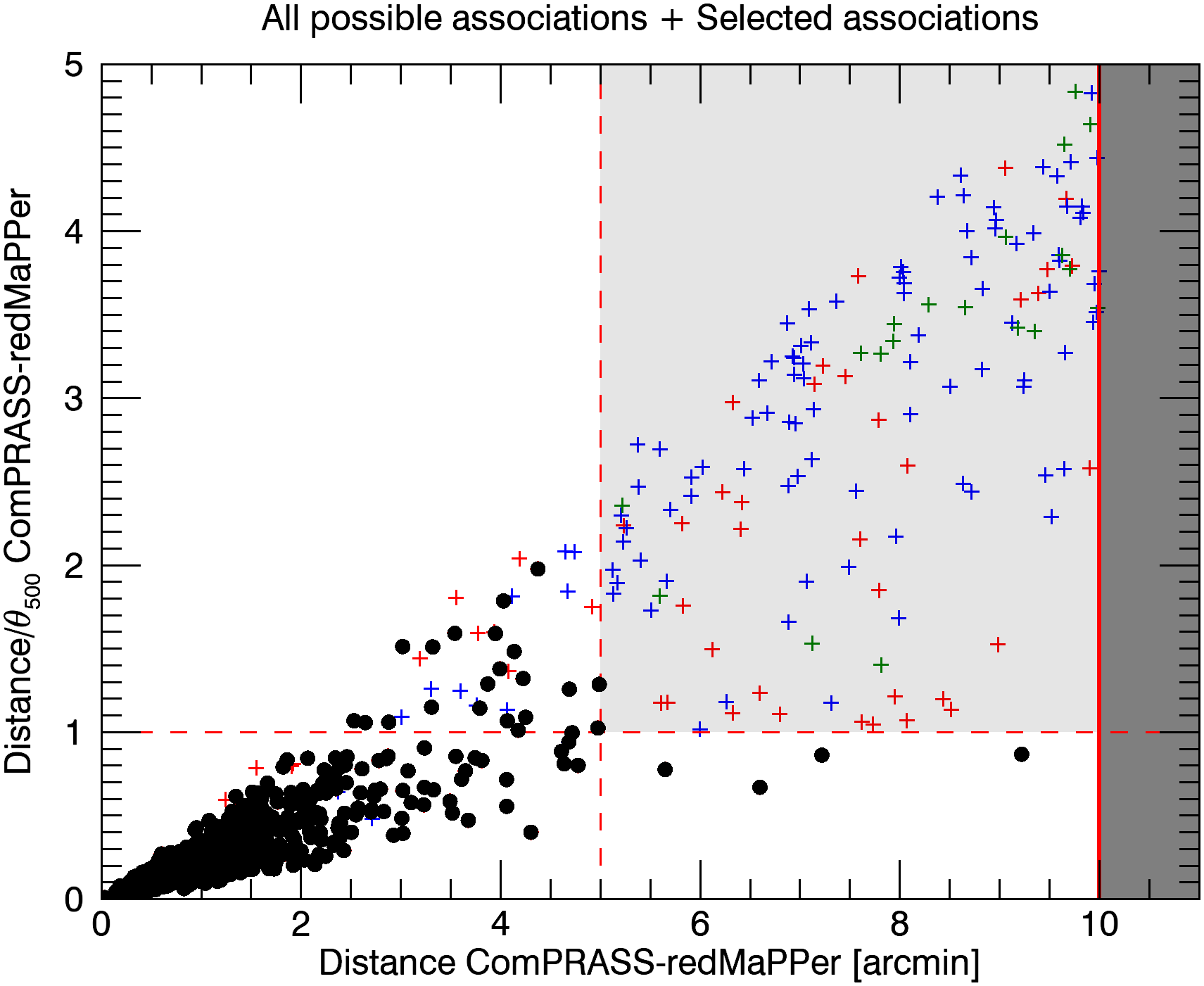}\label{fig:matching_redm_zoom}}
	\caption{Positional criteria for matching ComPRASS candidates to redMaPPer clusters. Fig. \ref{fig:matching_redm_loglog} shows the closest redMaPPer cluster to each ComPRASS candidate. Fig. \ref{fig:matching_redm_zoom} shows all the possible redMaPPer counterparts within 10 arcmin of each ComPRASS candidate. Red, blue and green symbols represent the closest, 2nd-closest and 3rd-closest counterpart respectively, if it exists. The black circles identify the pairs that are finally selected. The blue and red color crosses in the $d<5$ region correspond to associations that satisfy the positional criteria, but that are finally discarded by the richness cut.}
	\label{fig:matching_redm}              
\end{figure*}

\begin{figure}[]
	\centering
	\includegraphics[width=0.99\columnwidth]{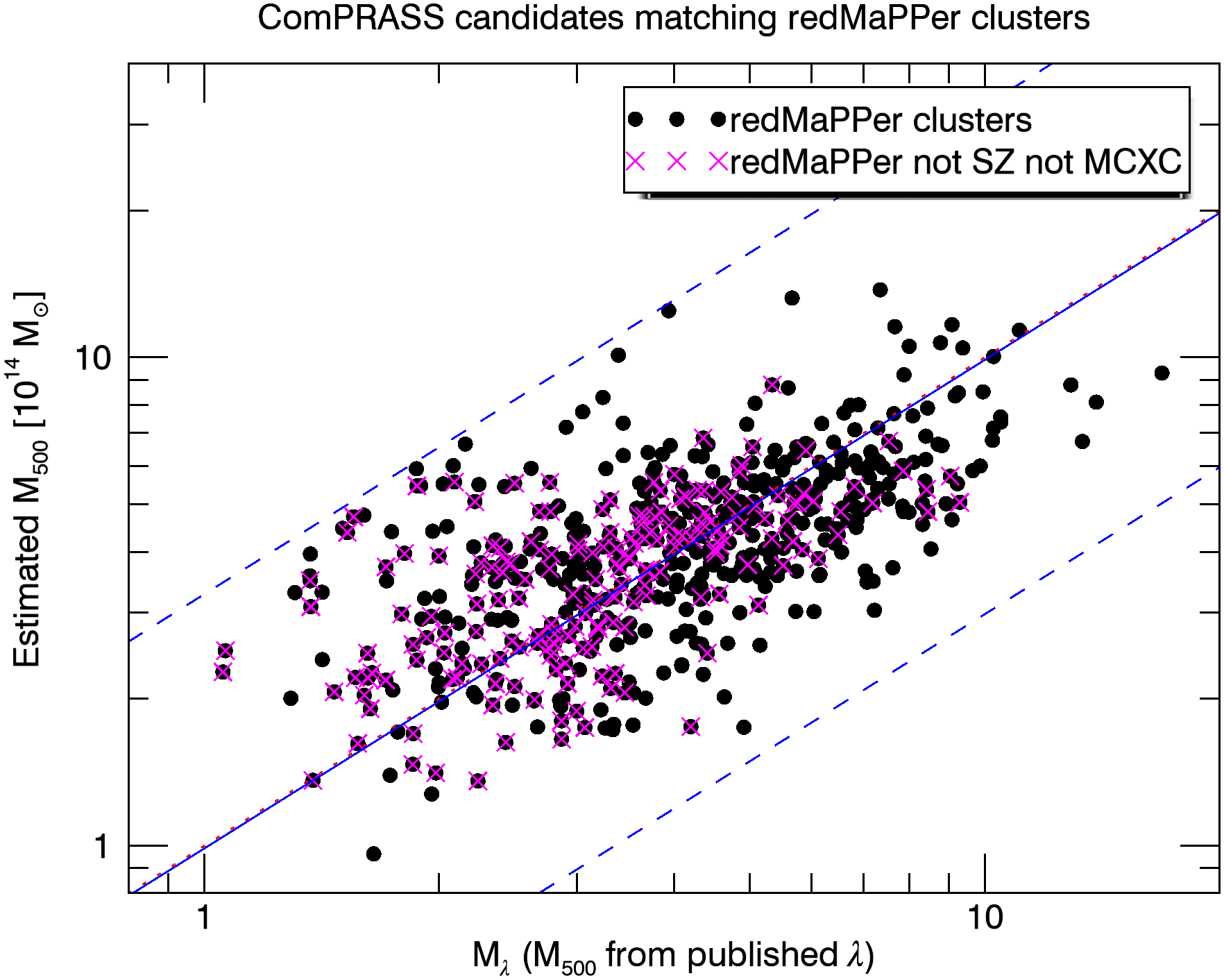}
	\caption{$M_{500}$ estimated from the joint detection for the 538 ComPRASS candidates matching a redMaPPer cluster vs. the $M_{500}$ of the corresponding cluster calculated from the richness (Eq. \ref{eq:mass-richness}). The dotted red line indicates the line of zero intercept and unity slope. The solid blue line indicates the median ratio. The dashed blue lines indicate the interval of $\pm 3\sigma$ around the median ratio. Magenta crosses indicate redMaPPer clusters that are not in \textit{Planck}, SPT, ACT or MCXC. }
	\label{fig:mass_xsz-mass_redMaPPer}              
\end{figure}

Regarding the eight 
candidates with underestimated mass, 5 of them are very extended and present an offset between the SZ and the X-ray peaks. This results in an underestimated mass because the extracted signal is not centered at the SZ peak. The three remaining candidates are probably clusters that are underluminous in X-ray, since the estimated mass using just the extracted SZ signal is perfectly compatible with the published SZ mass.

Finally, we also considered the catalogue of galaxy clusters presented in \cite{Burenin2017} detected in the \textit{Planck} all-sky Compton parameter maps and identified using data from the WISE and SDSS surveys. 
Since this catalog contains a large number of projections (about 37\%), we have decided to be conservative and consider these clusters as not yet confirmed SZ candidates (even though a big percentage of them will be real clusters). 
We found 604 matches at less than 10 arcmin distance from ComPRASS candidates. These candidates are classified as identified but not confirmed (class 2), unless they are confirmed from another catalogue. 
The information about these associations is included in the ComPRASS catalogue (see Appendix \ref{app:description}).

\subsection{Identification with optical clusters}\label{ssec:xmatch_opt}

We considered the seven optically selected cluster catalogues mentioned before: Abell, Zwicky, redMaPPer, MaxBCG, GMBCG, AMF, and  WHL. 
For the first two catalogues, there is no information readily available about the mass of the clusters, so we based our associations on positional criteria exclusively. The redMaPPer catalogue provides the richness of its clusters. The corresponding mass can be estimated using the richness-mass relation of \citet{Rozo2014}. We used this information, together with positional criteria to define our associations. For the four remaining SDSS-based catalogues, the richness is available, but the scatter of the richness-mass relation is too big, so we select our associations based directly on richness and positional criteria. In the rest of this Section, we describe in detail how the identification with the different catalogues was done.

\subsubsection{redMaPPer}\label{sssec:redMaPPer}

The redMaPPer catalogue contains clusters that were detected on SDSS data by looking for spatial over-densities of red-sequence galaxies. It provides an estimate of the richness and the photometric redshift for all the objects, and the spectroscopic redshift of the brightest central galaxy (BCG) for some of them. We used the latest version of the redMaPPer catalogue publicly available to date (v6.3), which contains 26111 objects.

The association of ComPRASS candidates with redMaPPer clusters was done in several steps, following a similar procedure to the one used by the \citet{Planck2015ResXXVII}, but with some differences. 

\begin{itemize}
	\item First, for each ComPRASS candidate, we look for all the redMaPPer clusters within a distance of 10 arcmin and we keep up to three redMaPPer clusters for each candidate. 
	\item For each possible association, we estimate the mass $M_{\rm J}$ of the candidate  by following the procedure described in Sect. \ref{ssec:catalog} assuming the redshift of the corresponding redMaPPer cluster. We used the spectroscopic redshift when available, otherwise we took the photometric redshift. This yields up to three masses for each candidate. 
	\item Then, as in the association with X-ray and SZ clusters, we discard all the pairs for which $5'<d<10'$ and $d/\theta_{500}>1$. In this case, $\theta_{500}$ is estimated from the richness and redshift reported in the redMaPPer catalogue, using the richness-SZ mass relation from \citet{Rozo2014}: 
	\begin{equation}\label{eq:mass-richness}
	\left\langle\rm{ln}\lambda|M_{\rm \lambda}\right\rangle = a + \alpha \rm{ln}\left(\frac{M_{\rm \lambda}}{M_{\rm p}}\right)
	\end{equation}
	with $a = 4.572 \pm 0.021$, $\alpha = 0.965 \pm 0.067$, $M_{\rm p} = 5.23 \cdot 10^{14}M_\odot$ and $\sigma_{\rm ln \lambda | M_{\lambda}} = 0.266 \pm 0.017$. This relation was calibrated with \textit{Planck} SZ masses, so we expect it to be well adapted to our validation.
	\item As done in the PSZ2 catalogue \citep{Planck2015ResXXVII}, we also discard the pairs for which the estimated mass of the candidate, $M_{\rm J}$, and the mass of the redMaPPer counterpart, $M_{\rm \lambda}$, calculated from the richness using Eq. \ref{eq:mass-richness}, differ significantly. In particular, we used the following rule: $|\rm {ln}(M_{\rm J}/M_{\rm \lambda})| > 3\sigma$, where $\sigma$=0.399 is the dispersion of $\rm {ln}(M_{\rm J}/M_{\rm \lambda})$. This value was calculated from the closest association to each candidate, after discarding those for which $5'<d<10'$ and $d/\theta_{500}>1$, and using an iterative procedure to discard the $3\sigma$ outliers.  
	We note that this value is higher than the one used by the \citet{Planck2015ResXXVII}, because we need to take into account both the dispersion of the richness-mass relation and the dispersion of our estimated mass $M_{\rm J}$ with respect to the true mass ($\sigma^2 = \sigma^2_{\rm{ln} \lambda|M_{\rm \lambda}}/\alpha^2 + \sigma^2_{M_{\rm J}|M_{\rm SZ}}$).
	\item After these two cuts in separation and mass, 538 candidates remain with an associated redMaPPer cluster, including 9 with 2 possible associations.
	\item Finally, if there is more than one possible association for a given candidate, we select the closest redMaPPer. We note that this choice is different from the one used in the PSZ2 catalogue, where the richest redMaPPer cluster was selected. This is justified because the position provided by the joint detection method is better than the one provided by \textit{Planck}. We also checked in detail these 9 cases and, taking into account the position and size of the SZ and X-ray peaks together with the position of the two possible redMaPPer counterparts and the SDSS data on these regions, we concluded that the closest redMaPPer is always a consistent match.
\end{itemize}

Fig. \ref{fig:matching_redm} shows the distribution of possible redMaPPer counterparts for ComPRASS candidates. Fig. \ref{fig:matching_redm_loglog} shows a scatter plot of the angular distance versus the relative distance between all the ComPRASS candidates and their closest redMaPPer counterpart. In this case, we do not observe clearly the two clouds corresponding to good and bad associations. Fig. \ref{fig:matching_redm_zoom} shows all the possible redMaPPer counterparts within 10 arcmin of each ComPRASS candidate. The black circles identify the pairs that are finally selected according to the above procedure (angular separation cut and richness cut). We see that the richness cut discards some of the associations that satisfy the angular separation criteria (blue and red color crosses in the left part ($d<5$ arcmin) of Fig. \ref{fig:matching_redm_zoom}).

Following the procedure described in Sect. \ref{ssec:catalog} and using the $M_{500} - D^2_A Y_{500}$ relation in Eq. \ref{eq:M500-Y500}, we estimated the mass $M_{500}$ for the 538 detections matching redMaPPer clusters. Figure \ref{fig:mass_xsz-mass_redMaPPer} shows the relation between the estimated mass and the mass of the corresponding clusters, calculated from the richess-mass relation of \citet{Rozo2014} (eq. \ref{eq:mass-richness}). This mass comparison shows that the ratio between the estimated mass and the published mass is very close to one, with a median value of 0.99.

\subsubsection{Abell and Zwicky}\label{sssec:abellzwicky}

Since there is no information about the mass of Abell \citep{Abell1989} and Zwicky \citep{Zwicky1961} clusters, and therefore about their size, we based our associations exclusively on the angular distance $d$ between the ComPRASS candidates and its closest Abell or Zwicky cluster. If this distance is lower than 5 arcmin, we consider that the ComPRASS candidate is associated to the Abell or Zwicky cluster. Indeed, taking into account their number, the chance association within this radius is negligible (e.g $<0.25\%$ for Abell clusters). Otherwise they are not associated. With this simple rule we find that 693 ComPRASS candidates are associated to an Abell cluster and 332 to a Zwicky cluster. These candidates are considered to be confirmed (class 1), since  Abell and Zwicky clusters are expected to be  real clusters, even if the spectroscopic redshift is not available. This information is included in the ComPRASS catalogue (Table \ref{table:catalogue}).
We note that 558 of the 693 candidates associated to an Abell cluster had already been confirmed with MCXC, SZ or redMaPPer, 36 had been identified with SZ or MCXC candidates, and 99 had not yet been identified. For the 332 candidates associated to a Zwicky cluster, 283 had already been confirmed with MCXC, SZ or redMaPPer, 11 had been identified with SZ candidates, and 38 had not yet been identified.

\subsubsection{Other SDSS-based catalogues}\label{sssec:SDSS}

Apart from redMaPPer, we have used four additional cluster catalogues based on SDSS data: (1) the MaxBCG catalogue (13823 objects, \cite{Koester2007}); (2) the GMBCG catalogue (55424 objects, \cite{Hao2010}); (3) the AMF catalogue (69173 objects, \cite{Szabo2011}); and (4) the WHL catalogue (132684 objects, \cite{Wen2012}). 

Unlike Abell and Zwicky, these catalogues contain many low-mass clusters, so we cannot directly associate our candidates to any SDSS cluster found nearby. Namely, the mass of the optical counterpart should be compatible with that of the X-SZ candidate, as in the case of redMaPPer (Sect. \ref{sssec:redMaPPer}). Each of the SDSS-based catalogues provides an estimated richness, which could be use to calculate a mass proxy using the richness-mass relation found in \cite{Planck2013ResXXIX}. However, the scatter in this relation is very big, so to be conservative we have decided to only consider the SDSS clusters with a richness above a certain threshold, as done in \cite{Planck2013ResXXIX}. 

To chose this richness threshold, we looked at all the potential associations between ComPRASS candidates (not yet confirmed with the previous validation) and SDSS clusters within an angular distance of 5 arcmin. Using optical images from PanStarrs, SDSS and WISE, together with available ancillary X-ray images from XMM-Newton and Swift, and the filtered maps obtained as output of our detection method, we have seen that all the potential associations with clusters with richness above 50 seem to be correct. Thus, the final criterion to associate a ComPRASS candidate with a SDSS cluster is an angular distance $d$ between the ComPRASS candidate and the SDSS cluster lower than 5 arcmin; and a cluster homogenized richness\footnote{We homogenized the richness estimates provided by the different SDSS-based catalogues to that of WHL, by applying the correcting factors found in \cite{Planck2013ResXXIX}:  1.52, 1.75, and 0.74 for MaxBCG, GMBCG, and AMF, respectively.} $\lambda \ge 50$. This gives 499 ComPRASS candidates associated with a cluster from the considered SDSS-based catalogues. They are classified as confirmed (class 1). We note that 483 of the 499 had already been confirmed with X-ray, SZ, redMaPPer, Abell or Zwicky catalogues, 9 had been identified with SZ candidates, and 7 had not yet been identified in with the previous catalogues.

\subsubsection{High-redshift optical-infrared catalogues}\label{sssec:others}
Since the above considered optical catalogues do not contain high-redshift clusters, we decided to consider three additional cluster catalogues which were specifically targeted at this type of objects. 

First, we considered the sample of 1959 massive clusters of galaxies in the redshift range of 0.7 < z < 1.0 presented in \cite{Wen2018}, which were found by searching around spectroscopically confirmed z > 0.7 LRGs in SDSS. We found two matches within a 5 arcmin radius with ComPRASS candidates that were not confirmed in the previous external validation: PSZRX G180.69+46.41 corresponds to Wen's J092829.4+410715 at z=0.8194 and had not been identified before; and PSZRX G271.62+61.69 corresponds to J115417.3+022124 at z=0.7118, and had been identified (but not confirmed) with one of the clusters of \cite{Burenin2017}. The joint masses assuming these redshifts are $M_{500}=(5.9 \pm 1.1) \cdot 10^{14} M_{\odot}$ and $M_{500}=(7.0 \pm 1.3) \cdot 10^{14} M_{\odot}$, respectively. The masses reported by \cite{Wen2018}, $M_{500}=7.37 \cdot 10^{14} M_{\odot}$ and $M_{500}=3.05 \cdot 10^{14} M_{\odot}$, respectively, do not have an associated error, but taking into account the possible uncertainties in the mass estimation at those high redshifts, they are compatible. 
By visually inspecting available ancillary images from WISE, PanSTARRS, SDSS, and \textit{Swift}, both appear to be good associations, so we have classified these two candidates as confirmed (class 1). 

We found 14 additional matches with already confirmed ComPRASS candidates: the redshift in Wen's catalogue is compatible (difference lower than 0.06) with the redshift of the other ComPRASS counterpart for six of them, and is higher (difference bigger than 0.19) for eight of the matches. Appendix \ref{app:multiple_associations_wen} includes detailed notes on these eight associations. They correspond to either chance associations or real counterparts that contribute to our total joint signal.

\begin{table*}
	\caption{Some interesting galaxy clusters found close to the new candidates (class 3) of the ComPRASS catalogue. The search was done in the NED and SIMBAD databases.}
	\label{table:candidates_unknown_matching_nedsimbad_cases2study}
	\centering 
	\tiny
	\begin{tabular}{c c | c c c}
		\hline
		\noalign{\smallskip}
		Id. &  ComPRASS  & Name & Redshift & Separation\\
		&  name      &      &          & [arcmin]\\
		\noalign{\smallskip}
		\hline
		\noalign{\smallskip}
		779	 & PSZRX G115.09+28.55 & PLCKESZ G115.12+28.56    	& 0.169	& 1.47	\\
		1307 & PSZRX G210.74+08.03 & PSZ1 G210.76+08.02       	& 0.296	& 1.56	\\
		1463	 & PSZRX G234.29+20.47 & PLCKESZ G234.2-20.5      	& 0.27	& 0.36	\\
		1556 & PSZRX G246.41+67.77 & ZwCl 4333                	& 0.08057	& 0.87	\\
		1579 & PSZRX G249.37+40.82 & RX J1020.5-0550 			& 0.404 & 1.25 \\
		1587  & PSZRX G249.99+24.23 & PSZ1 G250.02+24.15 		& 0.400 & 4.62 \\
		1638 & PSZRX G255.85+41.58 & SPT-CL J0438-4907        	& 0.24	& 0.72	\\
		2018	 & PSZRX G302.72+25.82 & Abell 3527-bis      		& 0.20	& 0.99	\\
	\end{tabular}
\end{table*}

Second, we considered the high-redshift clusters (0.7<z<1.5) from the Massive and Distant Clusters of WISE Survey (MaDCoWS) presented in \cite{Gonzalez2018}. We found 8 matches of the ComPRASS candidates with this catalogue within a radius of 5 arcmin, one of them corresponding to one of the identified but not confirmed candidates. It is candidate PSZRX G113.26+48.41, already identified with PSZ2 G113.27+48.39, and which is situated at 0.7 arcmin from MOO J1359+6725. Since the redshift and richness of MOO J1359+6725 is not available, our ComPRASS candidate remains not confirmed. The other 7 matches correspond to already confirmed (class 1) candidates. Two of them have a redshift that is higher than the redshift of the other candidate counterpart, and they are located at larger angular separation, so they may correspond to chance associations. Appendix \ref{app:multiple_associations_madcows} includes detailed notes on these two associations. The remaining 5 matches do not have a redshift in the MaDCoWS catalogue.

Finally, we took the 44 high-z (0.5<z<1.0) clusters presented in \cite{Buddendiek2015}, initially selected by cross-correlating the RASS faint and bright source catalogues with red galaxies from SDSS DR8, and then followed-up with optical telescopes. We found 12 matches of the ComPRASS candidates with this catalogue within a radius of 5 arcmin. Two of them correspond to still not identified candidates (class 3), and 10 to already confirmed candidates (class 1).

High redshift cluster ClG-J120958.9+495352 \citep{Buddendiek2015}, at z=0.902, is close to candidate PSZRX G139.37+65.89. The joint mass assuming this redshift is $M_{500}=5.64 \cdot 10^{14} M_{\odot}$, which is very close to the mass reported by \cite{Buddendiek2015} ($M_{500}=(5.3 \pm 1.5) \cdot 10^{14} M_{\odot}$). We have therefore associated these objects and classified our candidate as confirmed (class 1). 

Cluster ClG-J231520.6+090711, at z=0.725, is at less than 1 arcmin from candidate PSZRX G086.90+46.91. Although \citet{Buddendiek2015} does not provides the mass of this cluster, it seems that the association is correct, so we have classified our candidate as confirmed (class 1).

For the 10 matches with an already confirmed candidate, the redshift in Buddendiek's catalogue is compatible (difference lower than 0.06) with the redshift of the other ComPRASS counterpart in 6 of the cases, and is higher (with a difference bigger than 0.08) for 4 of the matches. We studied these four cases in detail (see notes in Appendix \ref{app:multiple_associations_buddendiek}) and we found that two of them seem to be correcly associated both to the Buddendiek's clusters and to the previously found counterparts (they are indeed the same object), and that the redshift provided by Buddendiek is more accurate. In the other 2 matches there is a superposition of two different clusters: the Buddendiek's cluster (closer to the detection) and a redMaPPer cluster (more distant). In these cases, the association with the higher-z Buddendiek cluster is confirmed, whereas the lower-z redMaPPer cluster may contribute to total signal or just be a chance association.

\begin{figure*}[]
	\centering
	\subfigure[]{\includegraphics[width=0.99\columnwidth]{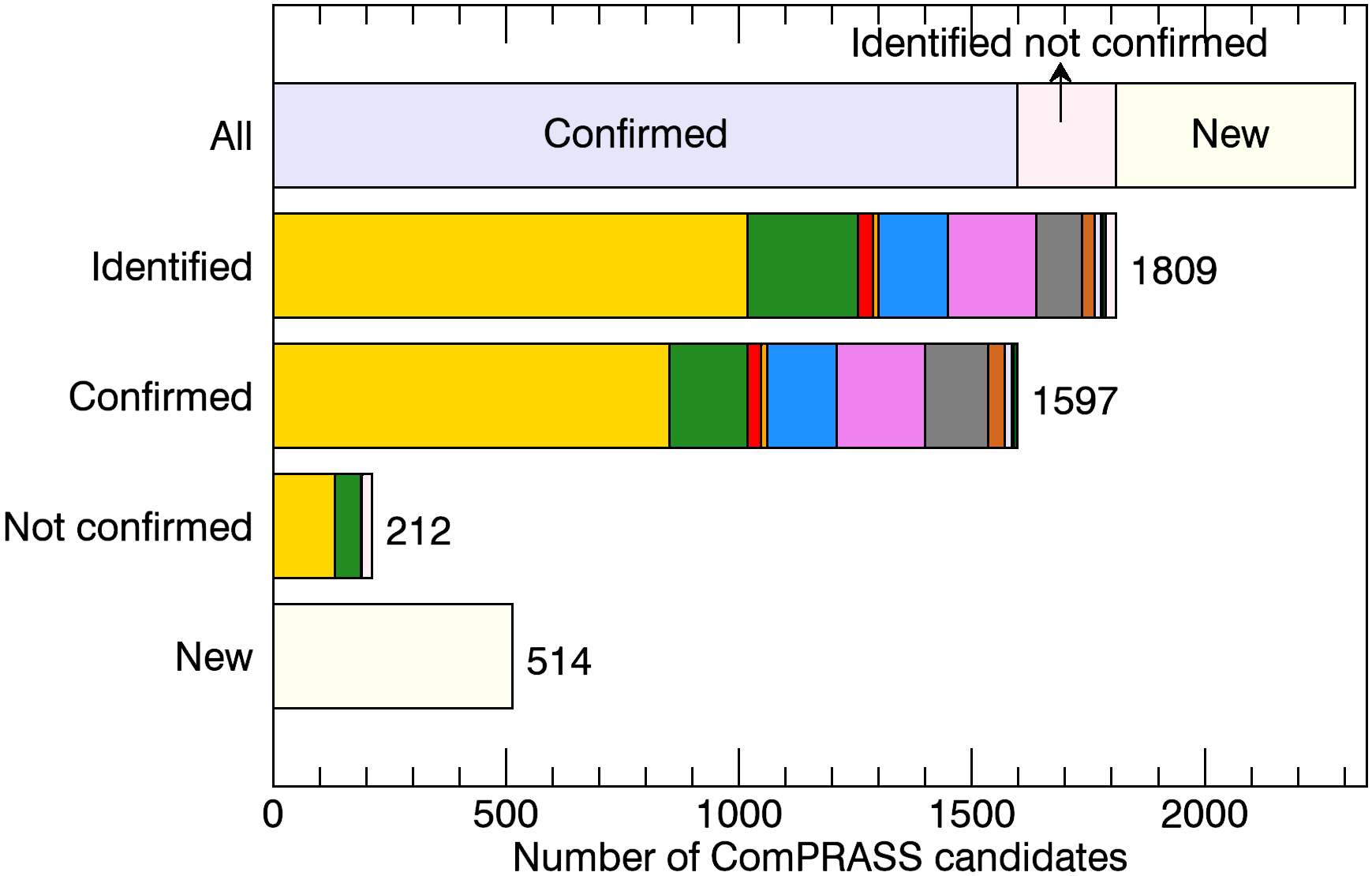}}
	\subfigure[]{\includegraphics[width=0.99\columnwidth]{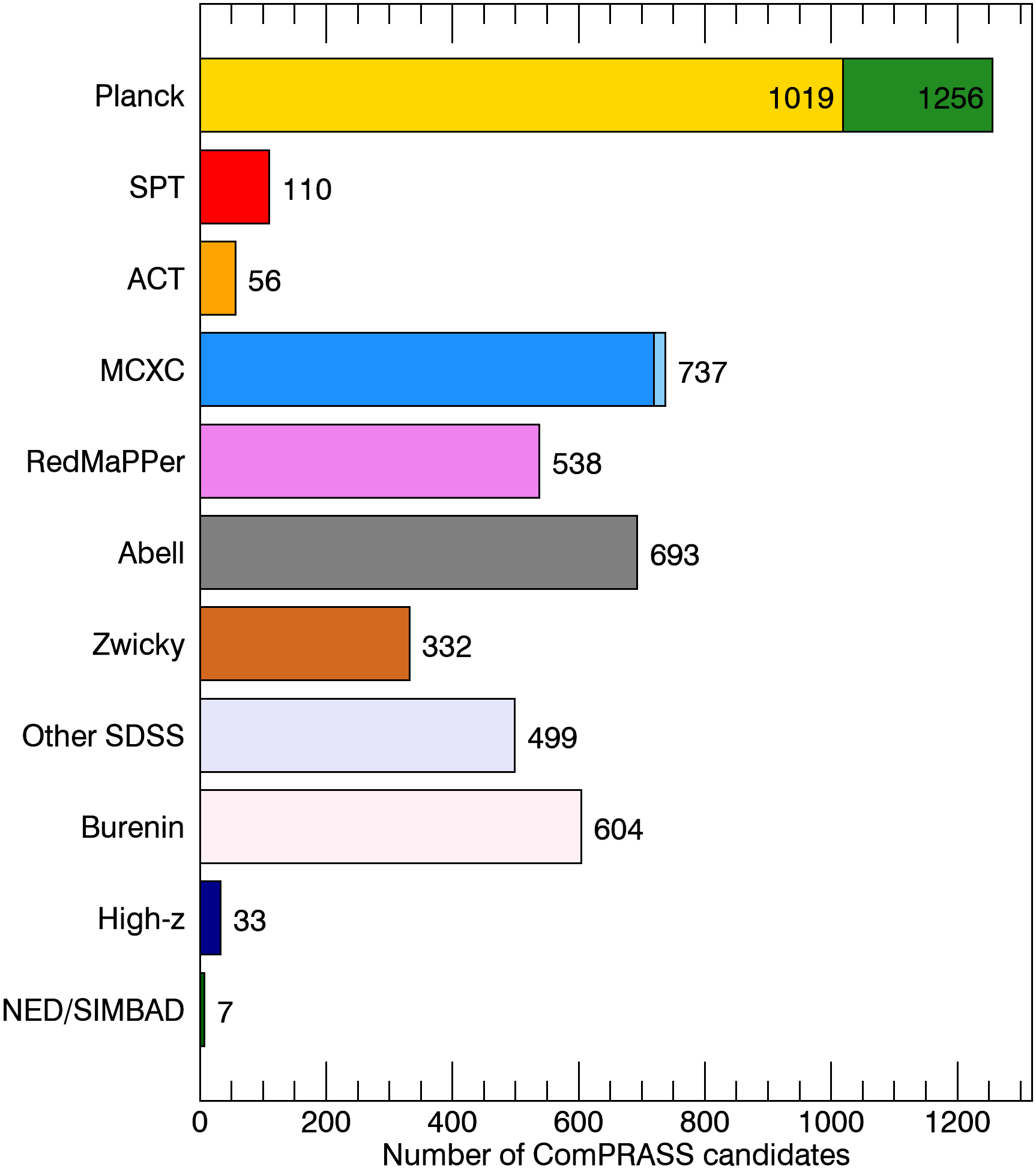}}
	\caption{(a) Number of candidates in the ComPRASS catalogue that were or not confirmed and/or identified with the external validation described in Sect. \ref{sec:validation}. \textit{Confirmed} (class 1) refers to the candidates that are associated to an already confirmed cluster; \textit{Identified not confirmed} (class 2) indicates the candidates that are associated to candidates in X-ray or SZ catalogues that have not been yet confirmed; and \textit{New} (class 3) represents the candidates that are not associated to any cluster or cluster candidate. \textit{Identified} includes candidates in class 1 and class 2. The different colors in the three middle bars correspond to the catalogue in which they were confirmed and/or identified: MMF3 (yellow), \textit{Planck} (green), SPT (red), ACT (orange), MCXC (blue), redMaPPer (violet), Abell (grey), Zwicky (brown), other SDSS-based catalogues (lavender), high-redshift catalogues (dark blue), additional clusters found in NED/SIMBAD (dark green), and \cite{Burenin2017} (pale pink). Since one candidate may have been identified with several catalogues, the indicated color refers to the first one in the previous list. For example, red indicates the candidates identified/confirmed by SPT and not identified/confirmed by \textit{Planck}. (b) Number of candidates in the ComPRASS catalogue that were identified with the different catalogues. This includes candidates classified as confirmed (class 1) and identified not confirmed (class 2). For the \textit{Planck} catalogues, the candidates identified with MMF3 are indicated in yellow, while the others are shown in green. For the MCXC catalogue, RASS and serendipitous clusters are indicated in medium and light blue respectively.}
	\label{fig:barplot_detected_clusters}
\end{figure*}

\begin{table*}
	\caption{Number of previously known clusters or cluster candidates that are associated to the ComPRASS candidates. Planck refers to the combination of the three \textit{Planck} catalogues \citep{PlanckEarlyVIII,Planck2013ResXXIX,Planck2015ResXXVII}, whereas PSZ2 refers only to the last one. MMF3 is the subsample of objects in the PSZ2 catalogue that were detected using the MMF3 detection algorithm. RASS refers to the subsample of objects in the MCXC catalogue that were detected from RASS observations. SZ refers to the combination of all the SZ catalogues (\textit{Planck}, SPT, and ACT).}
	\label{table:detectedclusters}
	\centering 
	\begin{tabular}{c c | c c }
		\hline
		\noalign{\smallskip}
		Cluster      &   Clusters in the   & Clusters  & Percentage \\
		catalogue    &   considered region & detected  &  (\%)      \\
		\noalign{\smallskip}
		\hline
		\noalign{\smallskip}
		all MMF3          &  1202 &  1019 & 84.8   \\   
		confirmed MMF3    &   909 &   850 & 93.5   \\
		MMF3 candidates   &   293 &   169 & 57.7   \\      
		\noalign{\smallskip}
		all PSZ2          &  1554 &  1200 & 77.2  \\             
		\noalign{\smallskip}
		all \textit{Planck}        & 1810 &  1256 & 69.4   \\           
		\noalign{\smallskip} 
		all SPT               &   647 &   110 & 17.0   \\
		ACT               &   206 &    56 & 27.2   \\
		\noalign{\smallskip}
		MCXC              &  1751 &   737 & 42.1  \\
		RASS              &  1294 &   719 & 55.6   \\
		Serendipitous      &   457 &    18 &  3.9  \\
		\noalign{\smallskip}
		redMaPPer             & 25333 &   538 &  2.1 \\
		Abell                 &  4995 &   693 & 13.9 \\
		Zwicky                &  8873 &   332 &  3.7 \\
		\hline
		\noalign{\smallskip}
		MCXC not SZ                 &  1122 &   150 & 13.4 \\
		\noalign{\smallskip}        
		all SPT not \textit{Planck} 			&   558 &    31 &  5.6 \\
		\noalign{\smallskip}
		all SZ                      &  2522 &  1299 & 51.5 \\            
		\noalign{\smallskip}
		all SZ + MCXC               &   3643 &  1449 & 39.8 \\          
		\noalign{\smallskip}
		\hline
		\noalign{\smallskip}
		\multicolumn{2}{c}{Total number of detections}            & \multicolumn{2}{|c}{2323}   \\
		\multicolumn{2}{c}{Detections matching any cluster candidate}     & \multicolumn{2}{|c}{1809} \\ 
		\multicolumn{2}{c}{New detections}     & \multicolumn{2}{|c}{514} \\ 
		\multicolumn{2}{c}{Detections matching a confirmed cluster}     & \multicolumn{2}{|c}{1597} \\ 
		\multicolumn{2}{c}{Detections matching a unconfirmed candidate}     & \multicolumn{2}{|c}{212} \\ 
		\hline
	\end{tabular}
\end{table*}

\subsection{Cross-identification with NED and SIMBAD databases}\label{ssec:nedsimbad}

Finally, we used NED\footnote{http://nedwww.ipac.caltech.edu/} and SIMBAD\footnote{http://simbad.u-strasbg.fr/simbad} databases to avoid missing a few additional associations. In particular, we looked for other known galaxy clusters around the ComPRASS candidates that have not been yet identified with the previous validation. Considering a search radius of 5 arcmin we found possible cluster counterparts for 60 ComPRASS candidates. Since most of these objects are small-mass optical clusters, it is difficult to determine if they really are the counterparts of our candidates. We focus instead on the ones that could be clear missed associations or particularly interesting cases.  

In particular, we found four \textit{Planck} clusters, one SPT cluster, one Zwicky cluster, one Abell cluster, and one X-ray cluster. Table \ref{table:candidates_unknown_matching_nedsimbad_cases2study} summarizes these possible missed associations. We studied these cases individually to determine why we missed some of them and whether or not they are good associations.
 
PLCKESZ G115.12+28.56 was not included in the published \textit{Planck} catalogues, but it was confirmed with optical follow-up observations at the Canary Island observatories \citep{PlanckIntXXXVI2016}. The distance between this cluster and ComPRASS candidate  PSZRX G115.09+28.55 is 1.47 arcmin, which means that they are associated. Consequently, we have classified this candidate as confirmed. 

PSZ1 G210.76+08.02 was not included in the published PSZ1 catalogue because it was below the 4.5$\sigma$ limit. However, it was confirmed with optical observations from the  Russian-Turkish 1.5 m telescope as part of the follow-up programme of the \textit{Planck} collaboration \citep{PlanckIntXXVI2015}. By looking at the \textit{Swift} observation of this cluster and knowing that it is 1.56 arcmin away from ComPRASS candidate PSZRX G210.74+08.03, we can conclude that they are associated. We have therefore classified this candidate as confirmed. 

PLCKESZ G234.2-20.5 was not included in the published ESZ catalogue because it was detected with a S/N lower than 6, which was the limit for the published catalogue. However, it was confirmed with \textit{XMM-Newton} observations in \cite{PlanckIntI2012}.  By looking at the available \textit{XMM-Newton} and \textit{Swift} observations of this cluster and knowing that it is only 0.36 arcmin away from ComPRASS candidate PSZRX G234.29+20.47, we can conclude that they are the same cluster. Therefore, we have classified this candidate as confirmed. Considering the redshift of PLCKESZ G234.2-20.5 (z=0.27), the mass estimated for PSZRX G234.29+20.47 is $M_{500}=5.07 \cdot 10^{14} M_{\odot}$, which is compatible with the mass of PLCKESZ G234.2-20.5 reported in \cite{PlanckIntI2012} ($M_{500}= (4.5 \pm 0.1)  \cdot 10^{14} M_{\odot}$) in the terms considered in Fig. \ref{fig:mass_xsz-mass_sz}.

PSZ1 G250.02+24.15 was considered in Sect. \ref{ssec:xmatch_XR}, but not associated with candidate 1587 because it was one of the matches falling in the grey zone of Fig. \ref{fig:matching_SZ} (see notes on Appendix \ref{app:notes}).

SPT-CL J0438-4907 was not included in the published SPT catalogue of \cite{Bleem2015} because its significance was lower than 4.5, which was the limit for the published catalogue. However, it was detected at lower significance and confirmed with optical observations in \cite{Saro2015}. The presence of this cluster at only 0.72 arcmin from candidate PSZRX G255.85+41.58 and the \textit{Swift} observation of this cluster indicate that this candidate is a real cluster \citep{Tarrio2018}. Moreover, the mass estimated assuming the redshift of the SPT cluster is $M_{500}=(3.0 \pm 0.5) \cdot 10^{14} M_{\odot}$, very close to the mass published by \cite{Saro2015} for the SPT cluster ($M_{500}=(3.13 \pm 0.81) \cdot 10^{14} M_{\odot}$). Thus, we have classified this candidate as confirmed.

ZwCl 4333 was not associated to ComPRASS candidate PSZRX G246.41+67.77 because they are separated by 13.73 arcmin. However, the position of ZwCl 4333 in SIMBAD is not the same as in the Zwicky catalogue. With this updated position, ZwCl 4333 and PSZRX G246.41+67.77 are separated 0.87 arcmin, so they would be associated. We have therefore classified this candidate as confirmed. 

Abell 3527-bis is not a cluster in Abell's catalogue. It was discovered by \cite{deGasperin2017} and named Abell 3527-bis due to its proximity (20 arcmin) and similar redshift (z=0.20) to Abell 3527 cluster. Candidate PSZRX G302.72+25.82 is at less than 1 arcmin from this cluster. Furthermore, the mass estimated assuming the redshift of this cluster is $M_{500}=3.02 \cdot 10^{14} M_{\odot}$, very close to the mass published by \cite{deGasperin2017} for Abell 3527-bis ($M_{500}=(3.3 \pm 0.81) \cdot 10^{14} M_{\odot}$). Therefore, we have decided to classify this candidate as confirmed (class 1). 

X-ray source RX J1020.5-0550 is galaxy cluster [ATZ98] C027 \citep{Appenzeller1998}, at redshift z=0.404. Candidate PSZRX G249.37+40.82 is only at 1.25 arcmin from this cluster and the detection peak coincides with the optical cluster. We have therefore classified this candidate as confirmed.

\subsection{Validation summary}\label{ssec:validation_results}

Fig. \ref{fig:barplot_detected_clusters} and Table \ref{table:detectedclusters} summarize the results of the identification of the ComPRASS candidates with the considered SZ, X-ray and optical cluster catalogues. From the 2323 candidates of the catalogue, 1597 correspond to known confirmed clusters, 212 are associated to a yet unconfirmed cluster candidate and 514 correspond to new objects. These 212+514 = 726 unconfirmed candidates could be either real clusters or false detections. Considering that the value of the purity estimated for the X-ray--SZ detection method in the SPT footprint ($>83.1\%$) \citep{Tarrio2018} is approximately valid for the whole sky, we expect to have at least 1930 real clusters in the catalogue, i.e. at least 333 real clusters among the 726 unconfirmed candidates.

\begin{figure*}[]
	\centering
	\includegraphics[width=1.9\columnwidth]{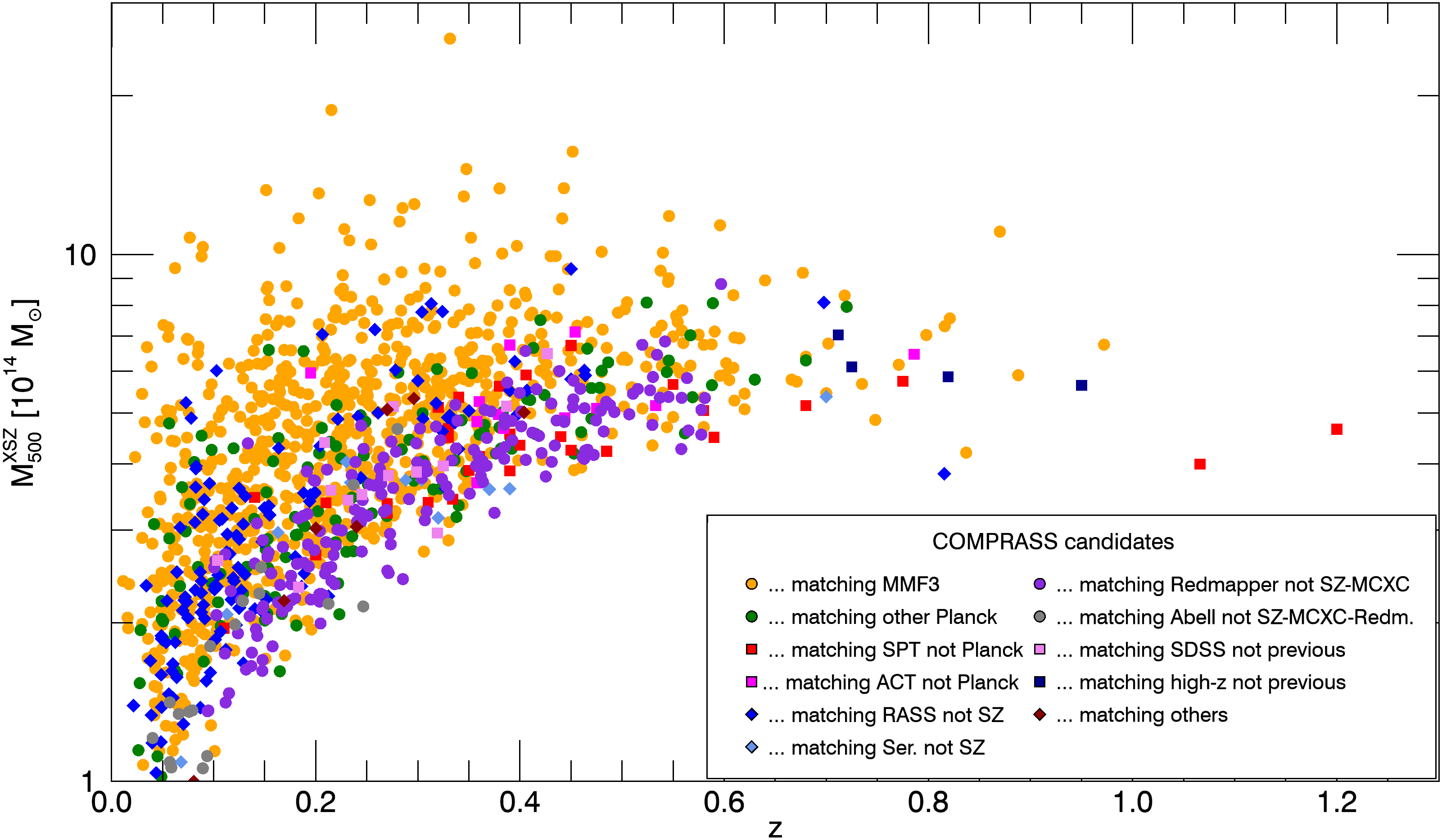}
	\caption{Distribution in the $M_{500}$--$z$ plane of validated ComPRASS candidates. Each candidate is color-coded according to its associated cluster. Yellow circles represent ComPRASS candidates matching confirmed MMF3 clusters, green circles represent ComPRASS candidates matching other confirmed \textit{Planck} clusters (not MMF3), red squares represent ComPRASS candidates matching confirmed SPT clusters not detected by \textit{Planck}, magenta squares represent ComPRASS candidates matching confirmed ACT clusters not detected by \textit{Planck} or SPT, blue diamonds represent ComPRASS candidates matching confirmed MCXC clusters that do not match any of the previously mentioned catalogues (dark blue for RASS, light blue for serendipitous), purple circles correspond to candidates matching redMaPPer clusters not in SZ or MCXC catalogues, grey circles correspond to candidates matching an Abell cluster not in the previous catalogues, pink squares correspond to candidates matching other SDSS clusters not in the previous catalogues, dark blue squares correspond to candidates matching other high-redshift clusters from \cite{Wen2018} and \cite{Buddendiek2015} not in the previous catalogues, and dark red diamonds correspond to candidates matching the additional clusters found in NED and SIMBAD databases. The mass is estimated from the X-SZ signal and the cluster $z$ as described in Sect. \ref{ssec:catalog}. 
	}
	\label{fig:detections_in_MZ_plane}
\end{figure*}

Fig. \ref{fig:detections_in_MZ_plane} shows	the distribution in the $M_{500}$--$z$ plane of the candidates in the ComPRASS catalogue that were validated from the external catalogues. The mass $M^{\rm XSZ}_{500}$ is the mass estimated by the X-ray--SZ detection method when the redshift of the associated cluster is assumed. Each cluster is color-coded according to the catalogue in which it was identified (see legend). If the cluster belongs to several catalogues, the following order is chosen: MMF3, other \textit{Planck}, SPT, ACT, MCXC, redMaPPer, Abell, other SDSS-based catalogues, and high-redshift catalogues. This figure evidences the ability to recover clusters that were not included in the MMF3 catalogue (yellow circles). In particular, at lower redshifts we gain lower-mass clusters discovered with X-rays, optical (redMaPPer), or even deeper SZ surveys (SPT and ACT). At high-redshift, we also recover some objects detected by deeper SZ surveys or deep X-ray observations. In particular, above z=0.75, we recover 1 cluster detected in the deep X-ray North Ecliptic Pole (NEP) survey \citep{Henry2006}, and 4 clusters detected by SPT and/or ACT. In comparison to MMF3, these additional clusters tend to populate the low mass regions at any redshift, pushing the detection limit of MMF3 towards a lower mass for any redshift.

\subsection{Redshift assembly}\label{ssec:redshift_priority}
The ComPRASS catalogue provides the redshift of all the candidates that have been confirmed by a cluster with an available redshift. If the ComPRASS candidate is associated to only one cluster, we provide the redshift of this cluster. In the case of multiple associations with different redshifts, we provide a unique redshift, which is chosen by applying the following order of priority: MCXC, SZ, redMaPPer (favoring spectroscopic over photometric redshift, when available), Abell, WHL, AMF, MAXBCG, GMBCG, and finally, the considered high redshift catalogues. We have privileged homogeneity for the sources of redshift rather than a case-by-case assembly of the most accurate redshift. We have checked that all the redshifts available for a given candidate are in general consistent. Some notes on specific cases were a discrepancy was found are included in Appendix \ref{app:multiple_associations}.

\section{Evaluation}\label{sec:evaluation}

 \subsection{Comparison with MMF3 clusters}\label{ssec:comparison_mmf3}
Since the X-ray--SZ detection method was built as an extension of the MMF3 detection method, we expect it to have a better performance than that of its predecessor. 

\begin{figure}[]
 	\centering
 	\includegraphics[width=0.99\columnwidth]{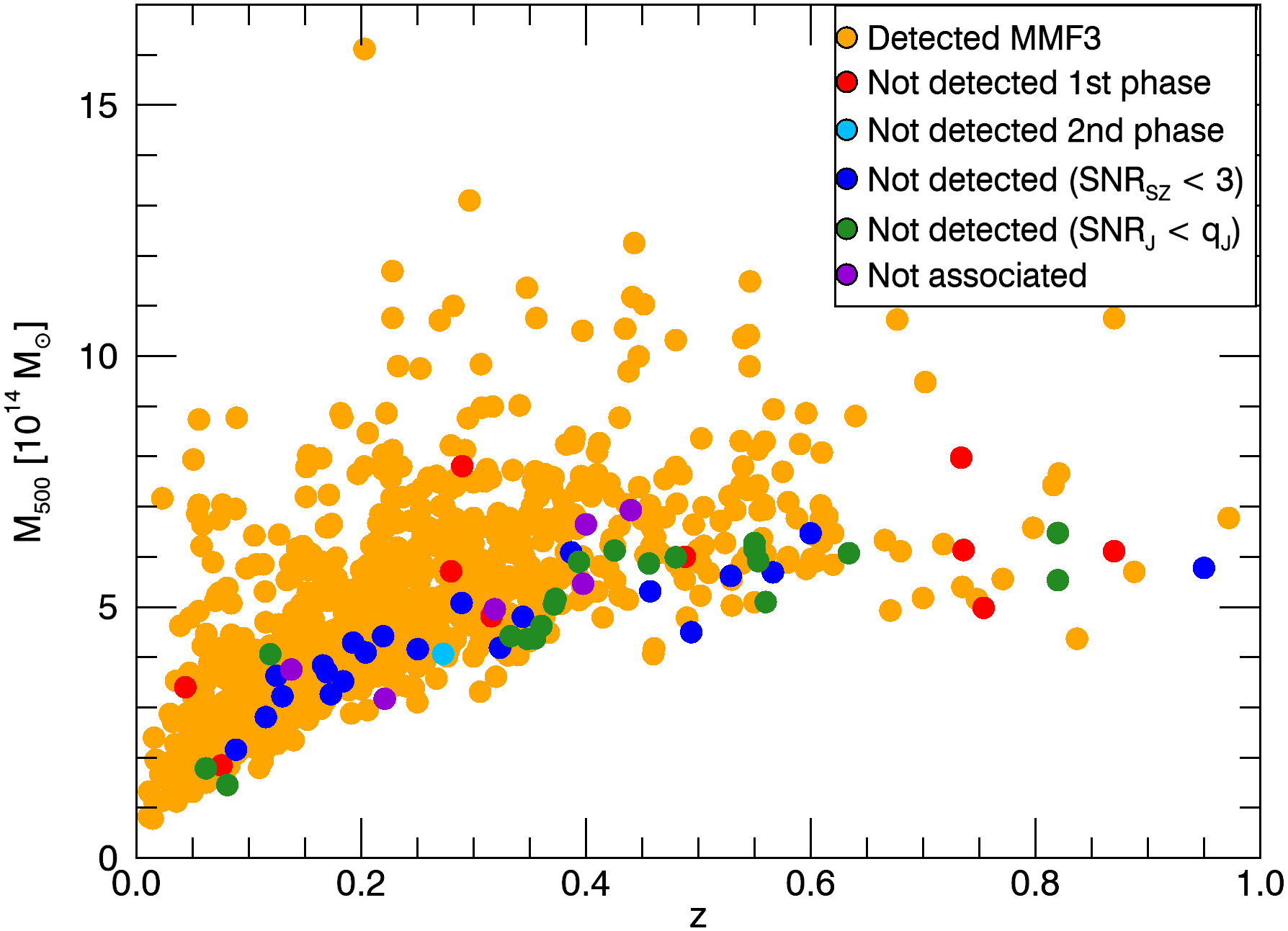}
 	\caption{Mass and redshift of confirmed clusters in the MMF3 catalogue. Yellow-filled circles represent MMF3 clusters that are detected by the joint detection algorithm; red-filled circles represent MMF3 clusters that were lost in the first phase of the algorithm; the light blue filled circle represents the MMF3 cluster that was lost in the second phase of the algorithm; dark blue filled circles represent MMF3 clusters detected in the second phase of the algorithm, but discarded due to a low (S/N)$_{\rm SZ}$; and green-filled circles represent MMF3 clusters detected in the second phase of the algorithm, but discarded because the joint S/N was not high enough.}
 	\label{fig:MMF3_clusters_in_MZ_plane}
 \end{figure}

As shown in Table \ref{table:detectedclusters}, the proposed method is able to recover 84.8\% of the MMF3 candidates (confirmed and not confirmed) included in the PSZ2 catalogue and situated in the considered region of the sky. There are 183 MMF3 candidates missing, 59 of which are confirmed clusters. Six of these 59 confirmed MMF3 clusters are detected within a radius between 5 and 10 arcmin but were not considered to be associated (grey zone in Fig. \ref{fig:matching_SZ_zoom}). 42 of the 59 missing clusters were initially detected (in the second phase), but then 20 were discarded because the joint S/N was not high enough and 22 were discarded because they had (S/N)$_{\rm SZ}$ < 3. One was detected in the first phase, but lost in the second phase because the joint S/N is below the initial threshold $q=4$. Finally, the remaining 10 were not detected in the first phase of the algorithm due to masking by another nearby detection (6) or due to a joint S/N below the initial threshold $q=4$ (4). Fig. \ref{fig:MMF3_clusters_in_MZ_plane} shows the distribution of the detected and missing MMF3 confirmed clusters in the mass-redshift plane.

The 124 MMF3 unconfirmed candidates that do not appear in our candidate list are missing for several reasons: 68 were initially detected in the second phase, but then 45 were discarded because the joint S/N was not high enough, 22 were discarded because they had (S/N)$_{\rm SZ}$ < 3 and 1 was discarded because it was in the masked region (survey mask). 10 were detected in the first phase, but lost in the second phase because the joint S/N is below the initial threshold $q=4$. Finally, the remaining 46 were not detected in the first phase of the algorithm due to masking by another nearby detection or due to a joint S/N below the initial threshold $q=4$.

There are in total 44 MMF3 objects (22 confirmed, 22 not confirmed) that we discard because (S/N)$_{\rm SZ}$ < 3. This may seem unexpected, since MMF3 objects have S/N>4.5 in the original extraction done by the \citet{Planck2015ResXXVII}. However, our value of (S/N)$_{\rm SZ}$ is measured at the position and size determined by the joint detection method, whereas the S/N value of original MMF3 objects is measured at the position and size determined by the MMF3 algorithm. Another difference with respect to MMF3 is that the resolution of the maps is different, since we are using a resolution of 0.86 arcmin/pixel instead of the original resolution of \textit{Planck} maps (1.72 arcmin/pixel). We used the MMF3 method of \cite{Planck2015ResXXVII} to extract the S/N of these 44 objects from \textit{Planck} maps in different ways. First, we centered the extraction at the position determined by the joint detection method, but we let the size as a free parameter. Second, we centered the extraction at the MMF3 position and we let the size as a free parameter. And finally, we used the maps with the original resolution (1.72 arcmin/pixel) and extracted the signal at the MMF3 position with free size. We found that 16 of the 44 MMF3 are lost due to the different size determined by the joint method, 17 are lost due to the different position determined by the joint method and 11 are lost because of the different resolution (plus the different position and size).

Even though the ComPRASS catalogue misses a small fraction of the MMF3 confirmed clusters (6.5\%), it includes other previously known clusters that are missed by MMF3 (see 
Table \ref{table:detectedclusters} and Fig. \ref{fig:detections_in_MZ_plane}). In particular it includes 168 additional \textit{Planck} clusters, 42 SPT and ACT clusters that were not detected by \textit{Planck}, 150 MCXC clusters that were not detected by \textit{Planck}, SPT or ACT, and 382 additional optical clusters not included in the considered SZ and X-ray catalogues. The overall effect is an improvement of the purity-detection efficiency performance with respect to the reference catalogue PSZ2-MMF3 (see also Fig. 13 of \citet{Tarrio2018}). 

Finally, it is interesting to understand the quality of the MMF3 detections that were missed or not by ComPRASS. The PSZ2 catalogue provides a quality flag, Q\_NEURAL, that indicates the quality of each detection. A value of Q\_NEURAL < 0.4 identifies low-reliability detections with a high degree of success \citep{Planck2015ResXXVII}. For the 1202 MMF3 candidates that fall in our unmasked region of the sky, the percentage of low-reliability detections is 4.4\%.  If we consider only the 1019 MMF3 candidates contained in the ComPRASS catalogue, the percentage of low-reliability detections falls down to 1.6\%, almost three times lower. The percentage of low-reliabilty candidates among the 183 MMF3 candidates that were missed by ComPRASS is 20.2\%. This indicates  the ability of the ComPRASS catalogue to clean the MMF3 catalogue from probably false detections, thanks to the incorporation of the X-ray information.

\subsection{Comparison with RASS clusters}\label{ssec:comparison_rass}
Since the proposed joint detection method uses RASS observations, it is interesting to check whether it is able to recover known clusters that have been  detected using the same observations. Table \ref{table:detectedclusters} shows that we detect 719 of the 1294 RASS clusters situated in the considered region (RASS exposure time greater than 100 s, outside the PSZ2 masked region), which corresponds to 55.6\%. There are 575 RASS cluster that we do not recover. Most of them (512) were in fact included in the list of detections provided by the second phase of the algorithm, but discarded for various reasons: 508 were discarded because their (S/N)$_{\rm SZ}$ was lower than 3, 2 were discarded because the RASS exposure time at the detection's position is lower than 100 s, 1 is discarded because it is outside the SZ mask and 1 was discarded because de joint S/N does not reach the adaptive threshold. 
 
There are 63 RASS clusters that were not originally detected by the joint algorithm. Most of them (49) were not detected because their joint S/N does not reach the threshold of $q=4$. Although these clusters are visible in RASS, the joint signal is not sufficiently high to be detected because their SZ signal is very faint: when we extract them from \textit{Planck} maps using the MMF3 method of \cite{Planck2015ResXXVII} they have a very low S/N, lower than 3 for 48 of them (so they would have been discarded anyway after the second phase of the algorithm) and 3.3 for the last one. The remaining 14 were masked by a nearby brighter source during the iterative peak detection procedure in the first phase of the algorithm. The clusters that are lost due to masking by close-by detections could be recovered in the future with an improved version of the iterative peak-finding process. Two possible solutions would be to use a smaller mask or to simultaneously search for all the peaks above a given threshold. In any case, a procedure to deblend sources would be needed afterwards.  

\begin{figure}[]
 	\centering
 	\includegraphics[width=0.99\columnwidth]{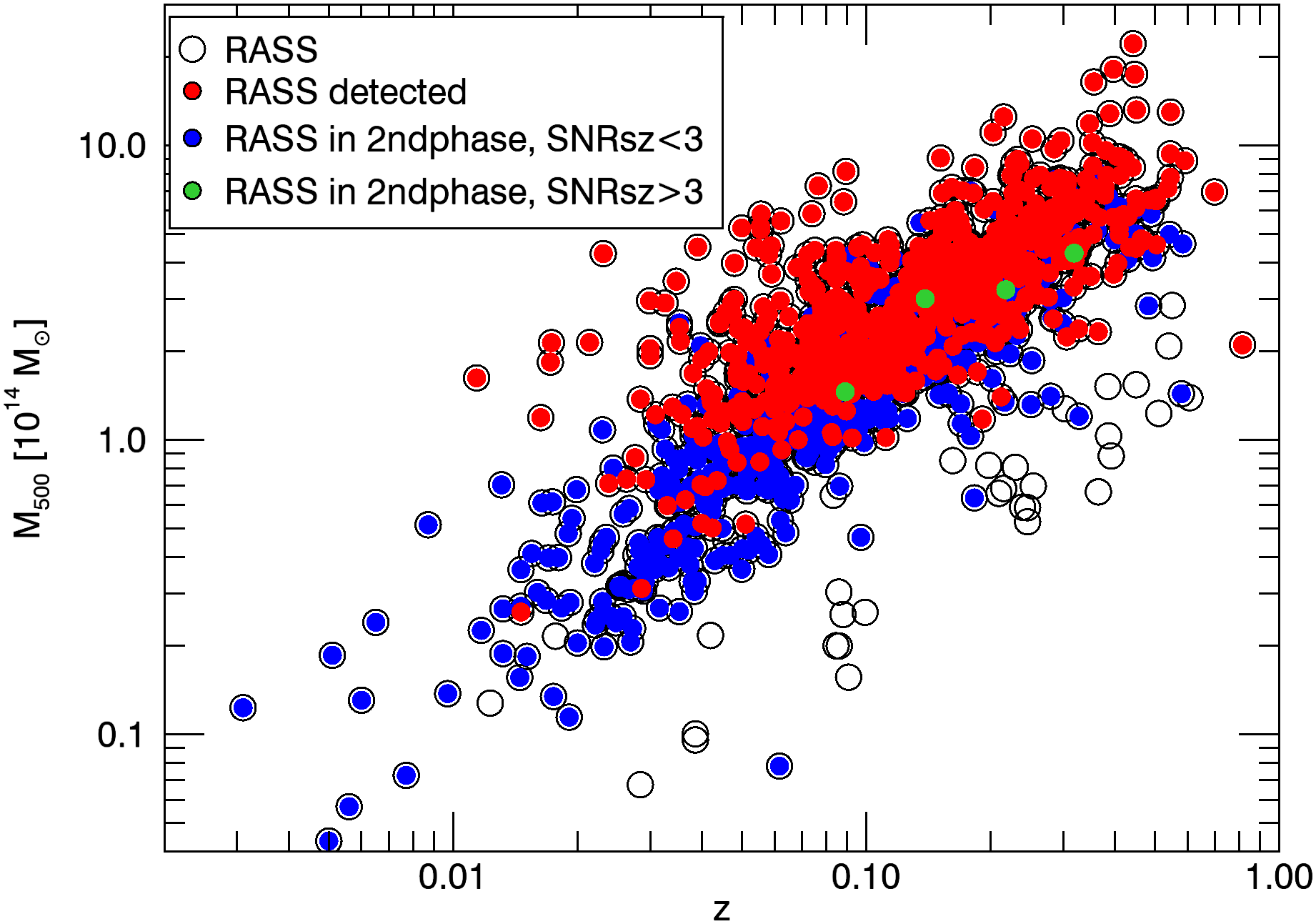}
 	\caption{Mass and redshift of the clusters in the RASS catalogue. Open circles represent RASS clusters in the considered region; red filled circles represent RASS clusters that are detected by the joint detection algorithm; blue filled circles represent RASS clusters that were detected in the second phase of the algorithm, but discarded due to a low (S/N)$_{\rm SZ}$; and the green-filled circles represent the RASS cluster that was detected in the second phase of the algorithm, but discarded due to other reasons (see text).}
 	\label{fig:RASS_clusters_in_MZ_plane}
 \end{figure}

In summary, the joint detection method is able to recover almost all the RASS clusters, as expected, but some of them are discarded later with the (S/N)$_{\rm SZ}$ threshold that we impose to remove AGN detections. Figure \ref{fig:RASS_clusters_in_MZ_plane} illustrates this comparison by showing the RASS clusters and the joint detections in the mass-redshift plane. The (S/N)$_{\rm SZ}$ threshold mainly eliminates low mass clusters that are detectable in RASS, especially at low redshift (blue filled circles). RASS clusters with very low mass at any redshift (empty circles) are mainly lost because their joint S/N does not reach the threshold of $q=4$, as explained before. 
 
The clusters of the MAssive Cluster Survey (MACS; \cite{Ebeling2001}, \cite{Ebeling2007}, \cite{Ebeling2010}, \cite{Repp2018}) are a particular subset of RASS clusters that were confirmed by a systematic optical follow-up of X-ray sources from the ROSAT bright source catalogues \citep{Voges1999,Boller2016}. Thus, they are clusters that are visible in the RASS maps, but which do not have a clear extended emission to be automatically detected as clusters. There are 109 MACS clusters in the considered region of the sky. The ComPRASS catalogue contains 75 of these clusters (68.8\%), whereas the PSZ2 catalogue contains only 55 (50.4\%). This shows that the addition of the X-ray information allows to detect more MACS clusters than those detected by \textit{Planck}. The 34 MACS clusters that do not appear in the ComPRASS catalogue are missing due to various reasons: 1 was masked by a nearby brighter source in the first phase of the algorithm, 2 were lost in the second phase because joint S/N did not reach the threshold of $q=4$, and the remaining 31 were detected in the second phase, but later discarded (1 because it was inside the SZ mask and 30 because their (S/N)$_{\rm SZ}$ was lower than 3).

 \subsection{Invalidated \textit{Planck} candidates}\label{ssec:unvalidated}
 \textit{Planck} published cluster catalogues contain confirmed clusters as well as not yet confirmed candidates that could correspond to false detections or to low-mass haloes boosted by SZ noise peaks. Since the publication of these catalogues, some of the candidates have been invalidated by specific follow-up observations. In particular, the MegaCam follow-up of \cite{vanderBurg2016} has invalidated 3 PSZ2 candidates and 8 PSZ1 candidates; and the optical follow-up observational programme developed at Roque de los Muchachos Observatory \citep{Barrena2018} has invalidated (labeled as 'Non detections', indicating that the counterpart of the SZ detection was not found) 49 PSZ1 candidates, two of which were already in the list of \cite{vanderBurg2016}. 
 
 The ComPRASS catalogue contains only five of these 58 invalidated objects: PSZ2 G037.67+15.71, PSZ2 G157.07-33.63, PSZ1 G029.79-17.37 (or PSZ2 G029.80-17.40), PSZ1 G044.83+10.02 (or PSZ2 G044.83+10.02) and PSZ1 G096.44-10.40. Since the percentage of \textit{Planck} candidates in the ComPRASS catalogue is 69.4\% and the percentage of invalidated \textit{Planck} candidates is only 8.6\%, this indicates that we tend to clean the \textit{Planck} catalogues from false candidates or low-mass systems. For this reason, we expect that the mass proxy provided by the ComPRASS catalogue will be less biased than the SZ mass proxy of the \textit{Planck} catalogues at low significance values.

 \subsection{Performance at high redshift}\label{ssec:highz}
 
  \begin{figure}[]
  	\centering
  	\includegraphics[width=0.99\columnwidth]{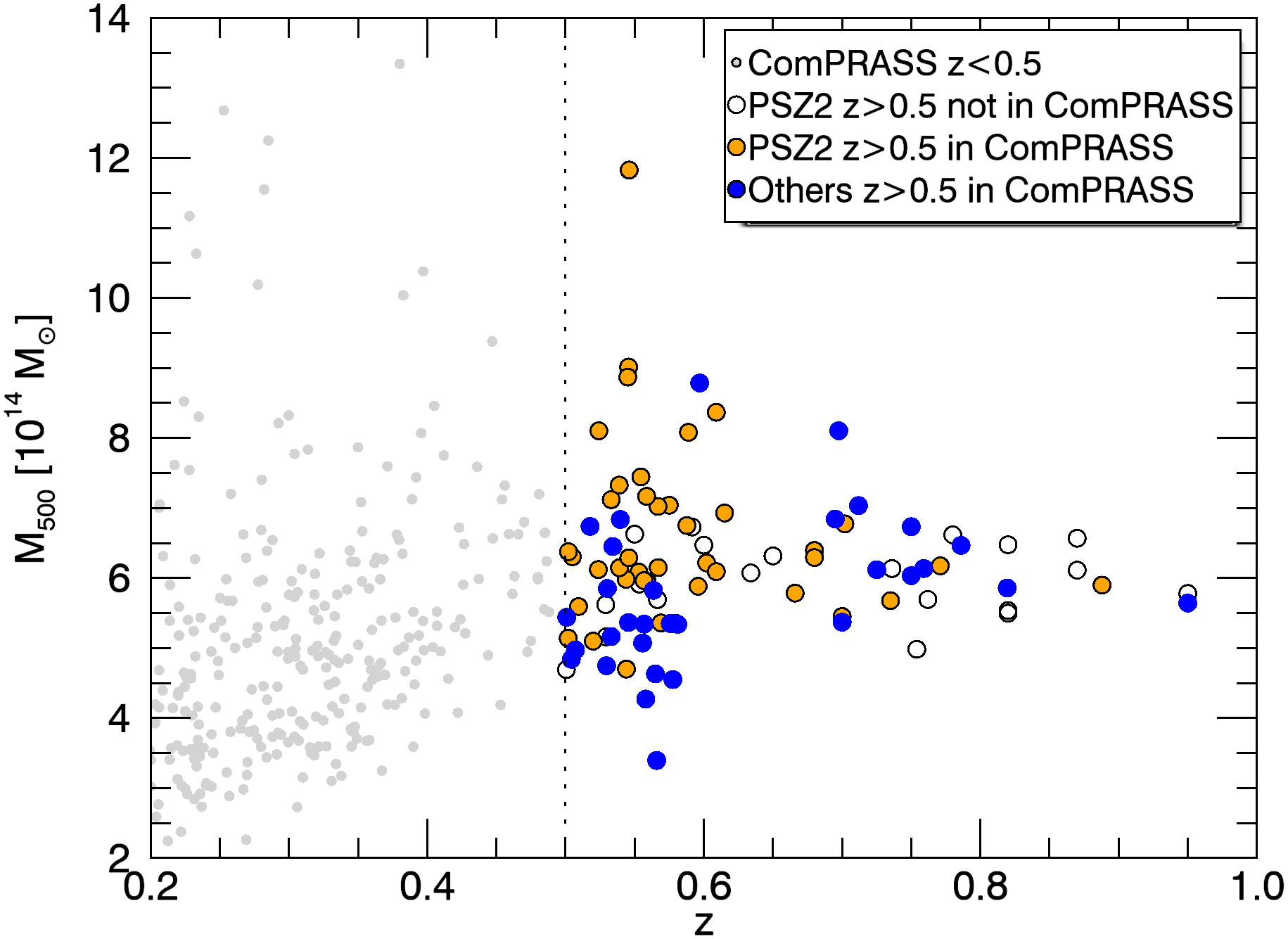}
  	\caption{Mass and redshift of known high-redshift (z>0.5) clusters in the SDSS footprint. Orange circles represent ComPRASS candidates that have been associated to a PSZ2 cluster at z>0.5. Blue circles represent ComPRASS candidates that have been associated to other z>0.5 clusters. Empty circles represent PSZ2 cluster at z>0.5 that are not included in the ComPRASS catalogue. Smaller grey circles show ComPRASS candidates confirmed at z<0.5. The mass represented in the figure corresponds to the joint mass estimated by the X-ray--SZ method for the grey, orange and blue circles; and to the published mass in the PSZ2 catalogue for the empty circles. }
  	\label{fig:detections_highz}
  \end{figure}
  
In this section we evaluate the performance of the ComPRASS catalogue in the high redshift regime (z>0.5). The catalogue includes 125 confirmed clusters in this regime, with more than half (73) located in the SDSS footprint, which covers 1/3 of the sky. The higher proportion of high-redshift clusters in the SDSS region is simply due to the fact that most of the optical catalogues considered during the validation cover only this area. We thus focus on this region to compare the performance of the ComPRASS catalogue to that of the PSZ2 catalogue. We remark, however, that SDSS-based catalogues are mostly limited to z<0.6, so we certainly have an incomplete list of high redshift ComPRASS clusters. 

Figure \ref{fig:detections_highz} shows the ComPRASS and PSZ2 high-redshift clusters in the mass-redshift plane. ComPRASS and PSZ2 contain 41 high-redshift clusters in common (orange circles). PSZ2 contains 20 high-z clusters that are not detected in ComPRASS (empty circles) and ComPRASS contains 32 high-z clusters not detected in PSZ2 (blue circles). The amount of high-redshift clusters that ComPRASS adds is a bit higher than the amount of high-redshift clusters that it looses with respect to the PSZ2 catalogue. However, due to the limited size of the sample and the lack of a systematic follow-up (some of the unconfirmed ComPRASS candidates could be indeed high-redshift clusters), it is difficult to extrapolate this behaviour to other regions of the sky.

 \subsection{Minimum expected purity}\label{ssec:sdssfootprint}

  \begin{table*}
  	\caption{Number and percentage of candidates that are classified in the three different classes (confirmed, identified not confirmed and new) in different regions: inside and outside the SDSS footprint; inside and outside the SPT footprint; inside and outside the SDSS and SPT footprints combined; and in all sky.
  		}
  	\label{table:sdssspt_footprint1}
  	\centering 

  	\begin{tabular}{c | c c | c c | c c | c c }
  	\hline
  	\noalign{\smallskip}
  	\noalign{\smallskip}
  	 & \multicolumn{4}{c|}{SDSS footprint} & \multicolumn{4}{c}{SPT footprint}   \\
  	 \noalign{\smallskip}
  	& \multicolumn{2}{c|}{Inside}  & \multicolumn{2}{c|}{Outside} & \multicolumn{2}{c|}{Inside}  & \multicolumn{2}{c}{Outside}  \\
  	\noalign{\smallskip}
  	&   Number & Percentage &   Number & Percentage &   Number & Percentage &   Number & Percentage \\
  	\noalign{\smallskip}
  	\hline
  	\noalign{\smallskip} 
  		Total          &   923  &        &  1400  &        &   225  &        &  2098  &        \\  
  		Confirmed      &   786  &  85.2  &   811  &  57.9  &   184  &  81.8  &  1413  &  67.3  \\  
  		Not confirmed  &    41  &   4.4  &   171  &  12.2  &     6  &   2.7  &   206  &   9.8  \\  
  		New            &    96  &  10.4  &   418  &  29.9  &    35  &  15.6  &   479  &  22.8  \\

  	\noalign{\smallskip}
    \hline
  	\end{tabular}
  	
  	 	\begin{tabular}{c | c c | c c | c c }
  	 		\hline
  	 		\noalign{\smallskip}
  	 		\noalign{\smallskip}
  	 		& \multicolumn{4}{c|}{SDSS/SPT footprint} & \multicolumn{2}{c}{All sky}   \\
  	 		\noalign{\smallskip}
  	 		& \multicolumn{2}{c|}{Inside}  & \multicolumn{2}{c|}{Outside} & \multicolumn{2}{c}{}   \\
  	 		\noalign{\smallskip}
  	 		&    Number & Percentage &   Number & Percentage &   Number & Percentage\\
  	 		\noalign{\smallskip}
  	 		\hline
  	 		\noalign{\smallskip} 
  	 		Total          &  1148  &        &  1175  &        &  2323  \\   
  	 		Confirmed      &   970  &  84.5  &   627  &  53.4  &  1597  &  68.7  \\  
  	 		Not confirmed  &    47  &   4.1  &   165  &  14.0  &   212  &   9.1  \\  
  	 		New            &   131  &  11.4  &   383  &  32.6  &   514  &  22.1  \\   
  	 		\noalign{\smallskip}
  	 		\hline
  	 	\end{tabular}

  \end{table*}

Given that the external validation was performed with cluster catalogues that do not cover homogeneously all the sky, it is interesting to see the differences between the regions of the sky that are better and less covered by the considered catalogues. Two particularly well covered regions are the SDSS footprint and the SPT footprint. All the clusters in the five considered SDSS-based catalogues (redMaPPer, MAXBCG, GMBCG, AMF, WHL) are contained in the former; whereas the SPT clusters are all located in the latter. This means that our validation is more exhaustive in these regions. 
 
Table \ref{table:sdssspt_footprint1} summarizes 
the number and percentage of candidates that are classified in the three different classes (confirmed, identified not confirmed, and new) in different regions: inside and outside the SDSS footprint, inside and outside the SPT footprint, inside and outside the two footprints combined, and in all the sky. 
The percentage of candidates that are classified as confirmed is higher in the SDSS and SPT regions (84.5\%) than in the rest of the sky (53.4\%). Since the chances that a ComPRASS candidate is not a real cluster should not depend on being or not in one of these regions, we expect to have at least the same percentage of real clusters in the rest of the sky as in the SDSS and SPT footprints. Therefore, with future validation efforts, we expect to find at least (1175*0.845)-627 = 365 real clusters among the 383+165 = 548 not confirmed and new candidates situated outside these two footprints.

\section{Conclusions}\label{sec:conclusions}

We have presented ComPRASS, the first all-sky catalogue of X-ray--SZ sources constructed by using a joint X-ray--SZ detection method \citep{Tarrio2018} on RASS and \textit{Planck} maps. This joint detection method can be seen as an evolution of the MMF3 detection method, one of the MMF methods used to detect clusters from \textit{Planck} observations, that incorporates X-ray observations to improve the detection performance.

The catalogue, which contains 2323 cluster candidates, has been validated by careful cross-identification of its candidates with existing X-ray, SZ and optical cluster catalogues. With this validation, we have classified the ComPRASS candidates into three classes: 1) confirmed (1597), which corresponds to the candidates that are associated to an already known cluster, 2) identified not confirmed (212), which includes the candidates associated to an already known candidate from SZ or X-ray catalogues that has not been yet confirmed as a real cluster, and 3) new (514), which contains the rest of the candidates that have not been associated to any known cluster or cluster candidate.

With respect to the reference catalogue PSZ2-MMF3, the ComPRASS catalogue is simultaneously more pure and more complete, since although it misses a small fraction of the MMF3 clusters (6.5\%), it includes many other 
additional clusters (747) that are missed by MMF3. In particular, it includes 382 optical clusters not included in purely SZ or purely X-ray catalogues. In addition, the ComPRASS catalogue contains a much smaller percentage of low-reliability detections than the MMF3 catalogue, and a much smaller percentage of invalidated \textit{Planck} candidates than the average recovery rate of \textit{Planck} candidates. This indicates that ComPRASS cleans the \textit{Planck} catalogues from false candidates (or low-mass systems). We thus expect to provide a less-biased mass proxy than the one given in the \textit{Planck} catalogues, especially at low significance values where the SZ mass proxy is more affected by Eddington bias \citep{vanderBurg2016}.

Regarding the X-ray catalogues, we recover 55.6\% of the RASS clusters. The rest, which are mostly at low mass and low redshift, are lost mainly due to the cut imposed to avoid AGN detections, but they could be incorporated by relaxing this constraint. Interestingly, the ComPRASS catalogue contains 68.8\% of the considered MACS clusters (more than in PSZ2). Since these clusters are not seen as extended sources in RASS, and they have been confirmed by systematically following-up tens of thousands of X-ray sources, the ComPRASS catalogue provides a way to find some of this clusters in a more direct way.

Considering the results in the regions of the sky where our validation is more exhaustive (SPT and SDSS footprints, where many more clusters are already known), the expected purity for the ComPRASS catalogue is greater than 84.5\%. This means that we expect to have more than 365 real clusters, unknown to date, among the new or yet-to-confirm candidates, especially among those situated outside these two footprints.
  
We are currently working on the validation of the ComPRASS candidates that have not been confirmed with the external validation presented here. To this end we are performing visual inspection and quantitative analysis on ancillary data from WISE, SDSS, PanSTARRS, \textit{XMM-Newton} and \textit{Swift}. The results of this study will be part of a future paper.

\begin{acknowledgements}
This research is based on observations obtained with \textit{Planck} (http://www.esa.int/Planck), an ESA science mission with instruments and contributions directly funded by ESA Member States, NASA, and Canada. This research has made use of the ROSAT all-sky survey data which have been processed at MPE. The authors acknowledge the use of the HEALPix package \citep{Gorski2005}. This research has made use of the SIMBAD database, operated at CDS, Strasbourg, France. This research has made use of the NASA/IPAC Extragalactic Database (NED), which is operated by the Jet Propulsion Laboratory, California Institute of Technology, under contract with the National Aeronautics and Space Administration. The research leading to these results has received funding from the European Research Council under the European Union's Seventh Framework Programme (FP7/2007-2013) / ERC grant agreement n$^{\circ}$ 340519. The authors would like to thank Iacopo Bartalucci for his help in the preparation of the \textit{Swift} images that were used to check some individual candidates. 
\end{acknowledgements}

\bibliographystyle{aa} % style aa.bst
\bibliography{mybiblio} % your references Yourfile.bib

\begin{appendix}

\section{Description of the ComPRASS catalogue}\label{app:description}

The ComPRASS catalogue contains 2323 candidates. For each candidate, the catalogue provides the following fields (see Table \ref{table:catalogue}). The median position error estimated from the distances to SPT positions is 54 arcsec \citep{Tarrio2018}.

\begin{sidewaystable*}
	\caption{Cluster candidates in the ComPRASS catalogue. The different fields are described in Appendix \ref{app:description}. This table is available in its entirety in a machine-readable form.}
	\label{table:catalogue}
	\tiny
	\begin{tabular}{c c c c c c c c c c c c c c}
		\hline
		\noalign{\smallskip}
		Id &  Name & raj2000  & dej2000 &  glon   & glat   &   SNR\_J & SNR\_SZ & SNR\_XR & Significance  & M500\_list & M500\_list\_err\_low & M500\_list\_err\_upp & z\_list \\
		  &    &    &   &      &      &(S/N)$_{\rm J}$ & (S/N)$_{\rm SZ}$ & (S/N)$_{\rm XR}$ &    &   &   &  &  \\
		\noalign{\smallskip}
		\hline
		\noalign{\smallskip}
   0 & PSZRX G000.06+45.19 & 229.157 &  -0.969 &   0.061 &  45.192 &     10.05 &   6.40 &   7.76 &   7.57 &   &   &  & \\ 
   1 & PSZRX G000.21+28.06 & 298.014 & -39.718 &   0.211 & -28.056 &      6.02 &   3.13 &   6.21 &   5.03 &   &   &  & \\ 
   2 & PSZRX G000.35+77.81 & 203.781 &  20.153 &   0.346 &  77.814 &      5.27 &   3.82 &   3.75 &   4.87 &   &   &  & \\ 
   3 & PSZRX G000.40+41.86 & 316.089 & -41.354 &   0.403 & -41.864 &     19.08 &   9.61 &  19.11 &  14.20 &   &   &  & \\ 
   4 & PSZRX G000.62+48.15 & 324.445 & -41.074 &   0.622 & -48.152 &      6.72 &   3.11 &   6.34 &   4.89 &   &   &  & \\ 
   5 & PSZRX G000.76+35.70 & 307.971 & -40.613 &   0.757 & -35.699 &     13.11 &   6.50 &  13.83 &  10.06 &   &   &  & \\ 
   6 & PSZRX G001.50+35.78 & 308.172 & -40.026 &   1.499 & -35.778 &      5.86 &   3.71 &   5.45 &   4.87 &   &   &  & \\ 
   7 & PSZRX G001.56+32.35 & 303.776 & -39.454 &   1.555 & -32.350 &      6.59 &   3.01 &   6.87 &   5.59 &   &   &  & \\ 
   8 & PSZRX G001.84+46.92 & 322.772 & -40.310 &   1.844 & -46.923 &     12.88 &   4.51 &  13.61 &   9.20 &   &   &  & \\ 
   9 & PSZRX G002.08+22.13 & 291.345 & -36.478 &   2.083 & -22.128 &      5.70 &   4.00 &   4.26 &   4.85 &   &   &  & \\ 
   10 & PSZRX G002.20+66.99 & 213.118 &  14.001 &   2.199 &  66.992 &     6.85 &   3.85 &   5.94 &   5.54 &   &   &  & \\ 
	\end{tabular}
	
	\begin{tabular}{c c c c c c c c c c c c}
		\noalign{\smallskip}
		\noalign{\smallskip}
		\hline
		\noalign{\smallskip}
		Id    & Class & z & z\_ref  & M500  & M500\_err\_low  & M500\_err\_upp & Planck & SPT & ACT & MCXC & redMaPPer \\
		  &    &   &    &  $M_{500}$ &   &   &   &   &   &   &   \\
		\noalign{\smallskip}
		\hline
		\noalign{\smallskip}
   0 & 1 & 0.120 & MCXC &   2.52 &   0.76 &   0.67 & PSZ2 G000.04+45.13 &  &  & RXC J1516.5-0056  &  \\ 
   1 & 3 &       &  &  -1 &  -1  &  -1  &  &  &  &  &  \\ 
   2 & 3 &       &  &  -1 &  -1  &  -1  &  &  &  &  &  \\ 
   3 & 1 & 0.165 & MCXC &   5.25 &   0.55 &   0.54 & PSZ2 G000.40-41.86 &  &  & RXC J2104.3-4120  &  \\ 
   4 & 3 &       &  &  -1 &  -1  &  -1  &  &  &  &  &  \\ 
   5 & 1 & 0.342 & MCXC &   6.53 &   0.83 &   0.79 & PSZ2 G000.77-35.69 & SPT-CLJ2031-4037 &  & RXC J2031.8-4037  &  \\ 
   6 & 1 &       &  &  -1 &  -1  &  -1  &  &  &  &  &  \\ 
   7 & 1 &       &  &  -1 &  -1  &  -1  &  &  &  &  &  \\ 
   8 & 1 & 0.421 & MCXC &   6.42 &   1.13 &   1.03 &  & SPT-CLJ2131-4019 &  & SMACSJ2131.1-4019   &  \\ 
   9 & 2 &       &  &  -1 &  -1  &  -1  & PSZ2 G002.04-22.15 &  &  &  &  \\ 
   10 & 1 & 0.140 & RedmapperSpec &   1.94 &   0.54 &   0.48 &  &  &  &  & RMJ141231.8+140041.1 \\

	\end{tabular}

	\begin{tabular}{c c c c c c c c c c c c c}
		\noalign{\smallskip}
		\noalign{\smallskip}
		\hline
		\noalign{\smallskip}
		Id  & Abell & Zwicky & MaxBCG & GMBCG & AMF & WHL & Wen2018 & Buddendiek2015 & Gonzalez2018 & id\_burenin & Other\_clusters & Notes \\
		\noalign{\smallskip}
		\hline
		\noalign{\smallskip}
   0 & ABELL 2051 &  &  & J229.18966-00.99227 &  & J151644.1-005809 &  &  &  &    -1 &  &  \\ 
   1 &  &  &  &  &  &  &  &  &  &    -1 &  &  \\ 
   2 &  &  &  &  &  &  &  &  &  &    -1 &  &  \\ 
   3 & ABELL 3739 &  &  &  &  &  &  &  &  &    -1 &  &  \\ 
   4 &  &  &  &  &  &  &  &  &  &    -1 &  &  \\ 
   5 &  &  &  &  &  &  &  &  &  &    -1 &  &  \\ 
   6 & ABELL 3688 &  &  &  &  &  &  &  &  &    -1 &  &  \\ 
   7 & ABELL 3671 &  &  &  &  &  &  &  &  &    -1 &  &  \\ 
   8 &  &  &  &  &  &  &  &  &  &    -1 &  &  \\ 
   9 &  &  &  &  &  &  &  &  &  &    -1 &  &  \\ 
   10 & ABELL 1875 &  & MAXBCG 5433 &  &  &  &  &  &  &  1868 &  &  \\

	\end{tabular}
	
\end{sidewaystable*}

\begin{itemize}

\item ID: Identifier of the candidate.

\item Name: Name of the candidate.

\item raj2000: Right ascension (J2000) in degrees.

\item dej2000: Declination (J2000) in degrees.

\item glon: Galactic longitude in degrees.

\item glat: Galactic latitude in degrees.

\item SNR\_J: Joint signal-to-noise ratio ((S/N)$_{\rm J}$) obtained with the best filter size.

\item SNR\_SZ: SZ component of the S/N: (S/N)$_{\rm SZ}$.

\item SNR\_XR: X-ray component of the S/N: (S/N)$_{\rm XR}$. 

\item Significance: Significance of the detection. It is defined in \citet{Tarrio2018} as the significance value in a Gaussian distribution corresponding to the probability that the detection is due to noise. 

\item M500\_list: Vector containing different values of joint mass, in units of 10$^{14}$ M$_\odot$,  corresponding to different values of redshift (given in z\_list). Together with z\_list, this vector provides the mass-redshift degeneracy curve of the candidate.

\item M500\_list\_err\_upp: Vector containing the upper bound of the 68\% confidence interval on the mass-redshift degeneracy curve, in units of 10$^{14}$ M$_\odot$.

\item M500\_list\_err\_low: Vector containing the lower bound of the 68\% confidence interval on the mass-redshift degeneracy curve, in units of 10$^{14}$ M$_\odot$.

\item z\_list: Vector containing the different values of redshift that correspond to the values in M\_list, M\_list\_err\_upp, and M\_list\_err\_low.

\item Class: Result from the classification of the candidate: 1 (confirmed), 2 (identified not confirmed), 3 (new).

\item z: Redshift of the candidate, which is taken from the cluster associated to the candidate. In case of multiple associations, this redshift is chosen as described in \ref{ssec:redshift_priority}. If no cluster with known redshift is associated to the candidate, the default value -1 is shown.

\item z\_ref: Indicates the origin of the redshift: MCXC, SZ, redMaPPerSpec, redMaPPerPhot, WHL, AMF, MAXBCG, GMBCG, WenHan2018, Buddendiek2015, MadCOWS or NED/SIMBAD.
    
\item M500: Estimation of the cluster joint mass ($M_{\rm J}$), in units of 10$^{14}$ M$_\odot$. This field is set to -1 when the redshift of the cluster is unknown and thus the mass cannot be calculated.

\item M500\_err\_upp: Upper bound of 68\% confidence interval on the cluster joint mass, in units of 10$^{14}$ M$_\odot$. This field is set to -1 when the redshift of the cluster is unknown and thus the mass cannot be calculated.

\item M500\_err\_low: Lower bound of 68\% confidence interval on the cluster joint mass, in units of 10$^{14}$ M$_\odot$. This field is set to -1 when the redshift of the cluster is unknown and thus the mass cannot be calculated. 
    
\item Planck: In case of association with a cluster/candidate in one of the \textit{Planck} catalogues, this field indicates the name of the \textit{Planck} cluster/candidate.

\item SPT: In case of association with a cluster in the SPT catalogue, this field indicates the name of the SPT cluster.

\item ACT: In case of association with a cluster in the ACT catalogue, this field indicates the name of the ACT cluster.

\item MCXC: In case of association with a cluster in the MCXC catalogue, this field indicates the name of the MCXC cluster.

\item redMaPPer: In case of association with a cluster in the redMaPPer catalogue, this field indicates the name of the redMaPPer cluster.

\item Abell: In case of association with a cluster in the Abell catalogue, this field indicates the name of the Abell cluster.

\item Zwicky: In case of association with a cluster in the Zwicky catalogue, this field indicates the name of the Zwicky cluster.

\item MaxBCG: In case of association with a cluster in the MAXBCG catalogue, this field indicates the name of the MAXBCG cluster.
 
\item GMBCG: In case of association with a cluster in the GMBCG catalogue, this field indicates the name of the GMBCG cluster.
 
\item AMF: In case of association with a cluster in the AMF catalogue, this field indicates the name of the AMF cluster.
 
\item WHL: In case of association with a cluster in the WHL catalogue, this field indicates the name of the WHL cluster.
 
\item Wen2018: In case of association with a cluster in the \cite{Wen2018} catalogue, this field indicates the name of the \cite{Wen2018} cluster.
  
\item Buddendiek2015: In case of association with a cluster in the \cite{Buddendiek2015} catalogue, this field indicates the name of the \cite{Buddendiek2015} cluster.
   
\item Gonzalez2018: In case of association with a cluster in the \cite{Gonzalez2018} catalogue, this field indicates the name of the \cite{Gonzalez2018} cluster.
  
\item id\_burenin: In case of association with a cluster in the catalogue of \cite{Burenin2017}, this field indicates the identifier of the cluster in this catalogue. Otherwise, it is set to -1.
  
\item Other\_clusters: In case of association with a cluster in table \ref{table:candidates_unknown_matching_nedsimbad_cases2study}, this field indicates the name of the cluster.
    
\item Notes: Notes on specific candidates.

\end{itemize}

\section{Notes on the association of individual candidates}\label{app:notes}

\subsection{Grey zone on X-ray associations} 
\textbf{Candidate 611}: This candidate is considered to be not associated with its closest MCXC cluster (RXC J0012.9-0853), since it is in the light grey zone of Fig. \ref{fig:matching_XR}. This detection is in a complex X-ray region with three very close emission peaks. The candidate is centered at one of the peaks, coinciding with PSZ2 G094.46-69.65, ACT-CL J0012.8-0855 and RMJ001257.7-085829.5. However, RXC J0012.9-0853 is centered on the peak just next to it. This indicates that the association is not correct.

\textbf{Candidate 675}: This candidate is considered to be not associated with its closest MCXC cluster (RXC J1540.1+6611), since it is in the light grey zone of Fig. \ref{fig:matching_XR}. This detection is inside the core of Abell 2125, which is a complex system. However, the position of RXC J1540.1+6611, a cluster from the 400SD catalogue \citep{Burenin2007}, is not on the core, but on LSBXE, a more diffuse X-ray emission to the southwest (see Fig. 4 of \cite{Miller2004}). This confirms the non-association. This detection also matches with Zwicky 7603. 

\textbf{Candidate 1327}: This candidate is considered to be not associated with its closest MCXC cluster (RXC J0347.4-2149), since it is in the light grey zone of Fig. \ref{fig:matching_XR}. This detection coincides with Abell 3138 and PSZ2 G215.19-49.65, but REFLEX II cluster RXC J0347.4-2149 is centered on a double structure just below A3168 \citep{Chon2012}. This confirms the non-association.

\textbf{Candidate 1568}: This candidate is considered to be not associated with its closest MCXC cluster (RXC J0956.4-1004), since it is in the light grey zone of Fig. \ref{fig:matching_XR}. The RASS image shows three X-ray extended emissions. The detection is centered on one of them, while RXC J0956.4-1004 corresponds to another one, which confirms the non-association. This detection also matches with Abell 901 and PSZ2 G247.97+33.52. The association with the PSZ2 cluster is considered correct, since the detection is in the same SZ peak as the PSZ2 cluster. In the PSZ2 catalogue, however, PSZ2 G247.97+33.52 is associated with RXC J0956.4-1004.

\subsection{Zone I on X-ray associations}
We define Zone I as the area in Fig. \ref{fig:matching_XR} where $d<5$ arcmin and $d/\theta_{500} > 1$. According to the X-ray association criteria defined in Sect. \ref{ssec:xmatch_XR}, candidate-cluster pairs in this region are considered to be associated.
 
\textbf{Candidate 957}: This candidate is considered to be associated with its closest MCXC cluster (RXC J1241.5+3250). A detailed look at the X-ray and SZ S/N maps shows that the SZ signal contributes with a double structure, which causes the detection to slightly move away from the RXC J1241.5+3250 position (situated between the two peaks), but staying inside the same X-ray emission region. Although the relative distance $d/\theta_{500}>1$, the angular distance is only 3 arcmin, so we can consider that the association is correct.

\textbf{Candidate 1991}: 
This candidate is considered to be associated with its closest MCXC cluster (RXC J0129.4-6432). A detailed look at the X-ray and SZ S/N maps shows that this candidate and RXC J0129.4-6432 are inside the same SZ peak and the same X-ray emission region, so we can consider that the association is correct.

\subsection{Zone II on X-ray associations}
We define Zone II as the area in Fig. \ref{fig:matching_XR} where $d>5$ arcmin and $d/\theta_{500} < 1$. According to the X-ray association criteria defined in Sect. \ref{ssec:xmatch_XR}, candidate-cluster pairs in this region are considered to be associated.
 
\textbf{Candidate 927}: 
This candidate has a very extended emission, both in X-ray and in SZ, corresponding to a very low redshift cluster (z=0.065). Although the candidate position is further than 5 arcmin away from the closest MCXC cluster (RXC J1200.3+5613), it is well inside the same X-ray emission peak. We can thus consider that the association is correct.
 
\textbf{Candidate 1472}: 
This candidate has a very extended emission, both in X-ray and in SZ. The candidate position is inside the same X-ray emission region as the MCXC cluster RXC J1144.6+1945, a very low redshift cluster at z=0.021. Therefore, we can consider that the association is correct.
 
\textbf{Candidate 1706}: 
The X-ray and SZ filtered maps of this candidate show two bright and extended X-ray sources in the same extended SZ peak. The candidate is centered at one of the X-ray sources, while the closest MCXC cluster (RXC J0627.2-5428) is situated between the two. This candidate corresponds to merging cluster A3395 \citep{Lakhchaura2011}. The candidate is centered at the SW emission region, while RXC J0627.2-5428 is centered closer to the NE region. Everything is part of the same system, so we can consider that the association is correct.

\textbf{Candidate 2106}: 
This candidate has a very extended emission, both in X-ray and in SZ. It corresponds to a very low redshift cluster (z=0.099). The candidate position is inside the same X-ray emission peak as the MCXC cluster RXC J2359.3-6042, so we can consider that the association is correct. 

\subsection{Grey zone on SZ associations}

\textbf{Candidate 435}:
There are two clusters around this candidate: 
Zwicky 8586 and PSZ1 G070.59-30.48. The PSZ1 position is in the center of a complex SZ structure that contains two SZ peaks. The candidate is situated at the stronger peak, thus, it seems to be correctly associated to PSZ1 G070.59-30.48. Furthermore, the candidate is closer to Zwicky 8586, to which PSZ1 G070.59-30.48 is associated in the PSZ1 catalogue.

\textbf{Candidate 742}:
The SZ S/N map around this candidate shows two SZ peaks. While the candidate is situated at one of the peaks, the closest SZ cluster (PSZ1 G266.19+19.06) is between the two, meaning that the PSZ1 detection covers both of them. Thus, we can consider that the association is correct.

\textbf{Candidate 830}:
A detailed look at the X-ray and SZ S/N maps around this cluster show that the candidate is inside the SZ emission centered at PSZ2 G121.13+49.64, but its position is biased towards a strong X-ray emission. In addition, the candidate is associated with a Zwicky cluster (Zwicky 5680), which is associated with PSZ2 G121.13+49.64 in the PSZ2 catalogue. For these reasons, we can consider that the association is correct. 

\textbf{Candidate 1197}:
This candidate is clearly associated with an MCXC cluster 
(RXC J0728.9+2935). Its closest SZ cluster (PSZ2 G189.23+20.55) was associated with the RXC J0728.9+2935 in the PSZ2 catalogue, thus, we consider that the association is correct. A detailed look at the X-ray and SZ S/N maps shows 1 peak in the X-ray map and 2 peaks in the SZ map, one coinciding with the X-ray peak.  The candidate and the MCXC cluster are situated at the X-ray peak, while the PSZ2 cluster is situated closer to the other peak. All of them seem to be part of the same (complex) structure. 

\textbf{Candidate 1300}:
This candidate is clearly associated with an MCXC cluster (RXC J0959.7+2223). Its closest SZ cluster (PSZ2 G210.01+50.85) was not associated with RXC J0959.7+2223 in the PSZ2 catalogue, thus, we consider that the association is not correct.

\textbf{Candidate 1587}:
The position of this candidate is shifted with respect to the SZ cluster PSZ2 G250.04+24.14 due to the presence of a bright X-ray source next to the cluster (but probably outside the cluster). Thus we can consider that the association is not correct.

\textbf{Candidate 1718}:
The SZ and X-ray S/N maps around this candidate show two strong X-ray emissions which appear as a single object in the SZ maps. The candidate is situated at the southern X-ray peak, closely matching one MCXC cluster (RXC J0330.0-5235) and one Abell cluster (Abell 3128), both at very low redshift. Its closest SZ cluster (PSZ2 G264.60-51.07, SPT-CLJ0330-5228, ACT-CL J0330-5227) is situated at the northern X-ray peak and has an estimated redshift of 0.44. Therefore, we can conclude that there are two different clusters and that the candidate corresponds to the lower-redshift cluster, so it should not be associated with the higher-redshift SZ cluster.

\textbf{Candidate 1732}:
This candidate is associated with Abell 3436. Its closest SZ cluster (PSZ1 G266.19+19.06) was not associated with Abell 3436 in the PSZ1 catalogue. However, a detailed look at the X-ray and SZ S/N maps show that the three objects belong to the same structure, so we can consider that the association is correct.

\textbf{Candidate 2094}:
This candidate is at the same position of Abell 3666 and MCXC cluster RXC J2016.2-8047. Its closest SZ cluster (PSZ2 G313.00-30.01) sits at another X-ray peak where there is another MCXC cluster and another Abell cluster. Thus, the assotiation of this candidate with PSZ2 G313.00-30.01 is not correct.

\section{Notes on multiple associations}\label{app:multiple_associations}

\textbf{Candidate 610}: Double cluster. This candidate is associated to a PSZ2 cluster at z=0.453 and to an MCXC and Abell cluster at z=0.225. The two clusters contribute to the detection.

\subsection{Multiple associations with \cite{Wen2018} clusters}\label{app:multiple_associations_wen}

\textbf{Candidate 497}: There are two different clusters within 5 arcmin. One is a PSZ2 cluster at z=0.04 and another one is a Wen cluster at z=0.81. The X-ray and SZ images correspond clearly to a very extended (very low redshift) cluster.
 
\textbf{Candidate 647}: There are two different clusters within 5 arcmin. One is a redMaPPer cluster at z=0.18 and another one is a Wen cluster at z=0.96. The redMaPPer cluster is much closer and the X-ray image corresponds better to this cluster. 

\textbf{Candidate 919}: There are two different clusters within 5 arcmin. One is a redMaPPer and Abell cluster at z=0.21 and another one is a Wen cluster at z=0.82. The redMaPPer cluster is much closer. The X-ray image shows the emission from the redMaPPer clusters.

\textbf{Candidate 953}: There are 3 different clusters within 5 arcmin. One is a PSZ2 cluster at z=0.36, another one is an Abell and redMaPPer cluster at z=0.28 and another one is a Wen cluster at z=0.70. The PSZ2 and the Abell-redMaPPer appear to be the most probable counterparts. 

\textbf{Candidate 1020}: There are two different clusters within 5 arcmin. One is a MCXC cluster at z=0.07 and another one is a Wen cluster at z=0.77. The X-ray and SZ images correspond clearly to a very extended (very low redshift) cluster.

\textbf{Candidate 1131}: There are two different clusters within 5 arcmin. One is a SZ-MCXC-redMaPPer cluster at z=0.19 and another one is a Wen cluster at z=0.86. The first one is the clear association by looking at Swift and XMM images. 

\textbf{Candidate 1260}: There are two different clusters within 5 arcmin. One is a redMaPPer cluster at z=0.57 and another one is a Wen cluster at z=0.76. Both of them have a mass of 2.5. The X-ray image does not allows us to discriminate which is the best position. We may be seeing a superposition of the two clusters. 

\textbf{Candidate 1308}: There are two different clusters within 5 arcmin. One is a redMaPPer cluster at z=0.41 and another one is a Wen cluster at z=0.71. The redMaPPer cluster is more massive and located at a smaller angular separation. There is another cluster at z around 0.2 visible within the 5 arcmin.

\subsection{Multiple associations with \cite{Gonzalez2018} clusters}\label{app:multiple_associations_madcows}

\textbf{Candidate 1143}: There are two different clusters within 5 arcmin. One is a cluster at z=0.21 that is included in MCXC, PSZ2, Abell and redMaPPer catalogues and that coincides in position with our candidate. The other one is a MADCowS cluster at z=1.12 further from our candidate (at 4 arcmin). The correct association is thus the lower-redshift cluster. The MADCowS cluster does not appear to contribute much to the total filtered signal.
 
\textbf{Candidate 1165}: There are two different clusters within 5 arcmin. One is a redMaPPer cluster at z=0.50 and the other one is a MADCowS cluster at z=0.89. The redMaPPer cluster is situated at the same position of our candidate (distance of 0.3 arcmin), while the MADCowS cluster is further away (at 3.6 arcmin). The correct association is thus the redMaPPer cluster. The MADCowS cluster does not appear to contribute much to the total filtered signal.

\subsection{Multiple associations with \cite{Buddendiek2015} clusters}\label{app:multiple_associations_buddendiek}
 
\textbf{Candidate 1121}: There is a redMaPPer cluster at z=0.60 and a \cite{Buddendiek2015} cluster at z=0.67 coinciding in position with the candidate and with the X-ray peak in RASS maps. Given that the redMaPPer redshift is photometric, it appears that both clusters are indeed the same object. Thus, the association is correct for both. Since the redshift provided by \cite{Buddendiek2015} is spectroscopic (and several spectroscopic redshifts consistent with it are also available in SDSS) we can consider that this last redshift is more precise.
 
\textbf{Candidate 1130}: There is a PSZ2 cluster at z=0.57 and a \cite{Buddendiek2015} cluster at z=0.70 coinciding in position with the candidate and with the X-ray peak in RASS maps. By looking at the SDSS images, both clusters seem to be the same object. Thus, the both associations are correct. The redshift provided by Buddendiek is compatible with the photometric reshifts available in SDSS, but the redshift of the PSZ2 cluster, which comes from a redMaPPer photometric redshift, is lower. This may indicate that it was underestimated.  
 
\textbf{Candidate 1232}: there is a superposition of two different clusters: the Buddendiek's cluster (closer to the detection) and a redMaPPer cluster (more distant). 
Candidate 1232 has a joint estimated mass of 5.55 assuming the redshift of the redMaPPer counterpart, while the redMaPPer mass is 2.77. If we assume the redshift of the Buddendiek's cluster, the joint mass would be 6.73, whereas the mass reported by \citep{Buddendiek2015} for this cluster is 9.8. It is possible that both clusters contribute to the detection, or that we have just detected the high-z cluster. In that case, the redMaPPer clusters would not be correctly associated.
 
\textbf{Candidate 1411}: there is a superposition of two different clusters: the Buddendiek's cluster (closer to the detection) and a redMaPPer cluster (more distant). Our candidate has a joint estimated mass of 3.97 assuming the redshift of the redMaPPer counterpart, while the redMaPPer mass is 1.81. It is possible that both clusters contribute to the detection, or that we have just detected the high-z cluster. In that case, the redMaPPer clusters would not be correctly associated.

\end{appendix}

\end{document}